\documentclass[fleqn,usenatbib]{mnras}

\usepackage{mathptmx}

\usepackage[T1]{fontenc}

\DeclareRobustCommand{\VAN}[3]{#2}
\let\VANthebibliography\thebibliography
\def\thebibliography{\DeclareRobustCommand{\VAN}[3]{##3}\VANthebibliography}

\usepackage{graphicx}	
\usepackage{amsmath}	
\usepackage{amssymb}	
\usepackage{subfigure}
\usepackage{color}

\title[SN 2015bf: flash-ionised signatures]{SN 2015bf: a fast declining type \uppercase\expandafter{\romannumeral2} supernova with flash-ionised signatures}

\author[Lin et al.]{
Han Lin,$^{1}$
Xiaofeng Wang,$^{1,2}$\thanks{E-mail: wang\_xf@mail.tsinghua.edu.cn}
Jujia Zhang,$^{3,4,5}$
Weili Lin,$^{1}$
Jun Mo,$^{1}$
Alexei V. Filippenko,$^{6,7}$
\newauthor
WeiKang Zheng,$^{6}$
Peter J. Brown,$^{8}$
Danfeng Xiang,$^{1}$
Fang Huang,$^{9}$
Yongzhi Cai,$^{1}$
Tianmeng Zhang,$^{10}$
Xue Li,$^{1}$
\newauthor
Liming Rui,$^{1}$
Xinghan Zhang,$^{1}$
Hanna Sai,$^{1}$
Xulin Zhao,$^{11}$
Melissa L. Graham,$^{12}$
I. Shivvers,$^{6}$
\newauthor
G. Halevi,$^{6}$
H. Yuk,$^{13}$
and Thomas G. Brink$^{6}$\\
$^{1}$Physics Department and Tsinghua Center for Astrophysics (THCA), Tsinghua University, Beijing, 100084, China\\
$^{2}$Beijing Planetarium, Beijing Academy of Sciences and Technology, Beijing, 100044\\
$^{3}$Yunnan Observatories, Chinese Academy of Sciences, Kunming 650011, China\\
$^{4}$Key Laboratory for the Structure and Evolution of Celestial Objects, Chinese Academy of Sciences, Kunming 650011, China\\
$^{5}$Center for Astronomical Mega-Science (CAS), 20A Datun Road, Chaoyang District, Beijing, 100012, China\\
$^{6}$Department of Astronomy, University of California, Berkeley, CA 94720-3411, USA\\
$^{7}$Miller Senior Fellow, Miller Institute for Basic Research in Science, University of California, Berkeley, CA 94720-3411, USA\\
$^{8}$George P. and Cynthia Woods Mitchell Institute for Fundamental Physics and Astronomy, Texas A\&M University; \\
Department of Physics and Astronomy, 4242, TAMU, College Station, TX 77843, USA\\
$^{9}$Department of Astronomy, Shanghai Jiao Tong University, Shanghai 200240, China\\
$^{10}$Key Laboratory of Optical Astronomy, National Astronomical Observatories, Chinese Academy of Sciences, Beijing 100012, China\\
$^{11}$School of Science, Tianjin University of Technology, Tianjin 300384, China\\
$^{12}$DiRAC Institute, Department of Astronomy, University of Washington, Box 351580, Seattle, WA 98195, USA\\
$^{13}$Homer L. Dodge Department of Physics and Astronomy, The University of Oklahoma, 440 W. Brooks St., Norman, OK 73019, USA
}

\date{Accepted XXX. Received YYY; in original form ZZZ}

\pubyear{}

\begin{document}
\label{firstpage}
\pagerange{\pageref{firstpage}--\pageref{lastpage}}
\maketitle


\begin{abstract}
We present optical and ultraviolet photometry, as well as optical spectra, for the type \uppercase\expandafter{\romannumeral2} supernova (SN) 2015bf. Our observations cover the phases from $\sim 2$ to $\sim 200$\,d after explosion. 
The first spectrum is characterised by a blue continuum with a blackbody temperature of $\sim 24,000$\,K and flash-ionised emission lines. 
After about one week, the spectra of SN 2015bf evolve like those of a regular SN \uppercase\expandafter{\romannumeral2}. From the luminosity of the narrow emission component of H$\alpha$, we deduce that the mass-loss rate is 
larger than $\sim 3.7\times10^{-3}\,{\rm M_\odot\,yr^{-1}}$. The disappearance of the flash features in the first week after explosion indicates that the circumstellar material is confined within $\sim 6 \times 10^{14}$\,cm. Thus, we suggest that the progenitor of SN 2015bf experienced violent mass loss shortly before the supernova explosion.
The multiband light curves show that SN 2015bf has a high peak luminosity with an absolute visual magnitude $M_V = -18.11 \pm 0.08$\,mag and a fast post-peak decline with a $V$-band decay of $1.22 \pm 0.09$\,mag within $\sim 50$\,d after maximum light. 
Moreover, the $R$-band tail luminosity of SN 2015bf is fainter than that of SNe~II with similar peak by 1--2\,mag, suggesting a small amount of ${\rm ^{56}Ni}$ ($\sim 0.009\,{\rm M_\odot}$) synthesised during the explosion. 
Such a low nickel mass indicates that the progenitor of SN 2015bf could be a super-asymptotic-giant-branch star that collapsed owing to electron capture.

\end{abstract}




\section{Introduction}

Type \uppercase\expandafter{\romannumeral2} supernovae (SNe~II), characterised by abundant hydrogen in their spectra \citep{1941PASP...53..224M}, belong to a class of SNe with diverse observed properties, with Type \uppercase\expandafter{\romannumeral2}P (plateau) and Type \uppercase\expandafter{\romannumeral2}L (linear) being the most common subtypes \citep{1979A&A....72..287B}. 
Although SNe~IIP and SNe~IIL have different light-curve shapes, whether they form two truly distinct classes is still debated \citep[e.g.,][]{2012ApJ...756L..30A, 2014ApJ...786...67A, 2014MNRAS.442..844F, 2014MNRAS.445..554F, 2015ApJ...799..208S}. 
Theoretical studies suggested that the mass of the H envelope can affect the observed properties of SNe~II \citep{1983Ap&SS..89...89L,1992SvAL...18...43B,1993ApJ...414..712P,2015ApJ...814...63M,2016MNRAS.455..423M}. \citet{2014ApJ...786...67A} and \citet{2017ApJ...850...90G}
showed that the progenitors of fast-declining SNe have a relatively lower-mass H envelope at the time of explosion. 
The spectra of SNe~IIn show prominent narrow or intermediate-width emission lines, which are produced by the interaction between SN ejecta and circumstellar material \citep[CSM;][]{1990MNRAS.244..269S, 1997ARA&A..35..309F}.

Recent observations show that narrow emission lines appear not only in SNe~IIn, but also in early-time spectra of other subtypes of SNe~II. 
\cite{2014Natur.509..471G} referred to spectra showing such emission lines as ``flash ionised (FI).'' Such FI emission lines are formed as a result of recombination of the outermost CSM, which has been ionised by high-energy photons created during SN shock breakout \citep{2014Natur.509..471G}. \cite{2016ApJ...818....3K} estimated that at least $18 \%$ of SNe~II show FI signatures in spectra taken at sufficiently early times, and \citet{2020arXiv200809986B} suggested an even larger fraction ($ >30 \% $ at $ 95 \% $ confidence level) of the SNe~II detected in the first year of the Zwicky Transient Facility (ZTF) survey.


It is commonly accepted that SNe~II are produced by core collapse and explosion of massive stars ($\geq 8 {\rm M_\odot}$;  e.g., \citealt{2009ARA&A..47...63S}). However, the final stages of evolution of massive stars immediately before exploding is still unclear and direct observations are absent. Flash spectra may provide one of the best ways to infer the CSM properties and hence the mass-loss history of the late-phase evolution of the progenitor stars. 
For example, \cite{2017NatPh..13..510Y} studied the case of SN 2013fs and proposed that the CSM was ejected during the final $\sim 1$\,yr prior to the explosion at an enhanced rate, $\sim 10^{-3}\,{\rm M_\odot\,yr^{-1}} $. Moreover, the observations of SN 2013fs and SN 2017eaw \citep{2019MNRAS.485.1990R} also indicated that pre-SN instabilities may be common among massive stars. 

As FI features are usually seen in very early-phase spectra and disappear quickly, detections and studies of such features require rapid observations of SNe~II within a few days after explosion; the sample with detections of FI features is thus very limited. In this paper, we present observations and analysis of another SN~II (SN~2015bf) with noticeable FI features in early-time spectra, though it was initially classified as a Type \uppercase\expandafter{\romannumeral2}n supernova.
In Section \ref{sec:observe}, we describe the observations and data reduction. The photometric and spectroscopic evolution of SN~2015bf are presented in Sections \ref{sec:phot} and \ref{sec:spec}, respectively. We discuss the mass-loss history, explosion parameters, and possible progenitor properties in \ref{sec:discussion}. Section \ref{sec:conclu} summarises our conclusions.

\section{OBSARVATIONS AND DATA REDUCTION} \label{sec:observe}


\subsection{Discovery} \label{subsec:dsicovery}
\begin{figure}
\includegraphics[width=\columnwidth]{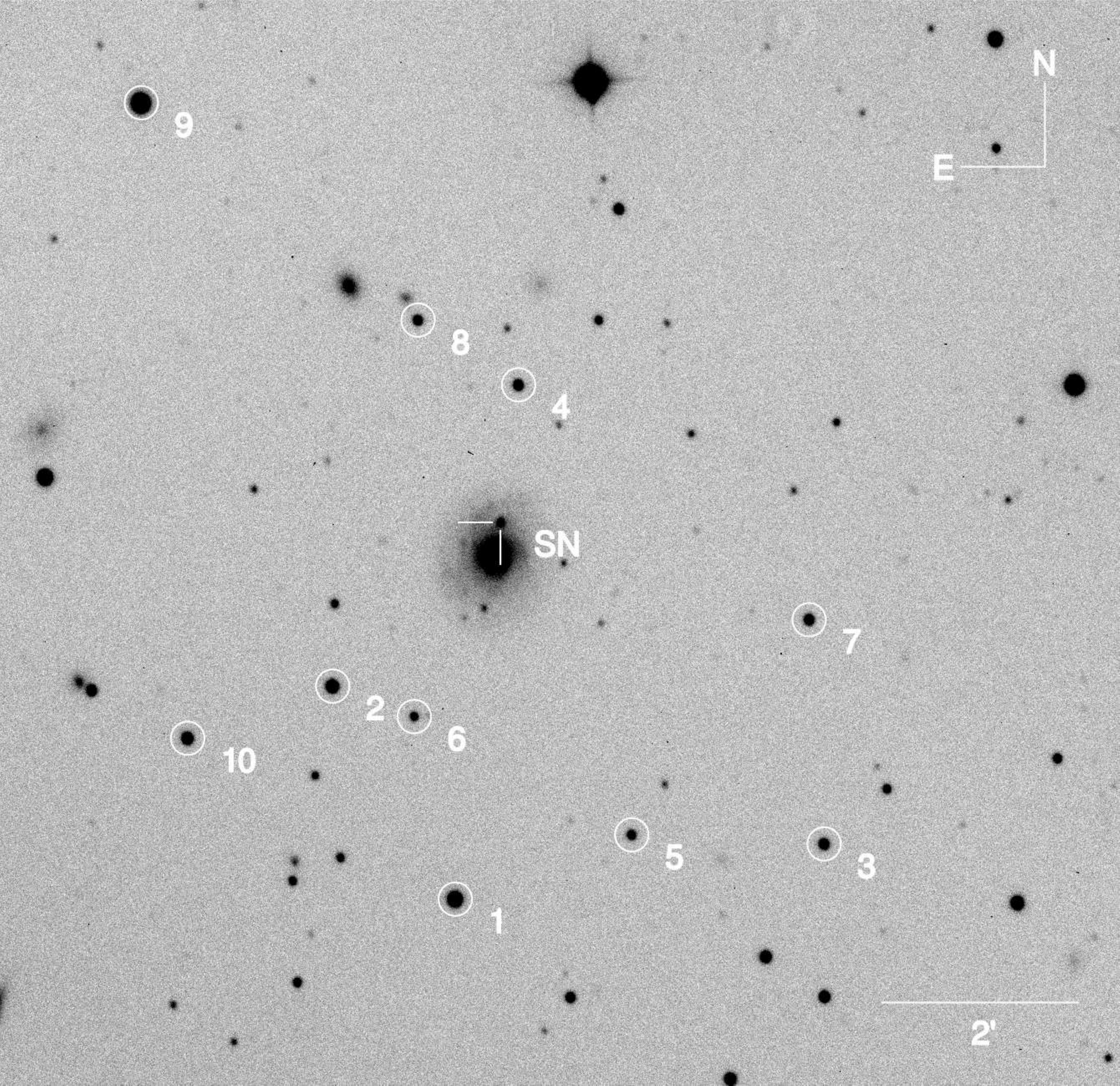}
\caption{$R$-band image of SN 2015bf in NGC 7653, taken on 2016 Feb. 1 with the Tsinghua-NAOC telescope. North is up and east is to the left. The locations of the supernova and local reference stars are marked by white tick marks and circles, respectively.  \label{fig:discovery}}
\end{figure}

SN 2015bf (= PSN J23244903+1516520) was discovered on 2015 December 12.468 (= JD 2,457,368.968; UT dates are used throughout this paper) 
by Koichi Itagaki using a 0.60\,m reflector, with an unfiltered magnitude of $\sim 17.3$\,mag. 
Its J2000 coordinates are $\alpha = 23^{\rm h}24^{\rm m}49.03^{\rm s}$ 
and $\delta = 15^{\circ}16^{\prime}52^{\prime\prime}.0$, located $4^{\prime\prime}.0$ west 
and $20^{\prime\prime}.0$ north of the centre of host-galaxy NGC 7653 (Fig. \ref{fig:discovery}). 
SN 2015bf was classified as a young Type \uppercase\expandafter{\romannumeral2}n SN from a spectrum taken on 2015 December 13.5, which showed narrow emission lines superimposed on a blue continuum \citep{2015ATel.8410....1C,2015ATel.8412....1Z}.
Extensive follow-up observations have been carried out of this object.

\begin{figure}
\includegraphics[width=\columnwidth]{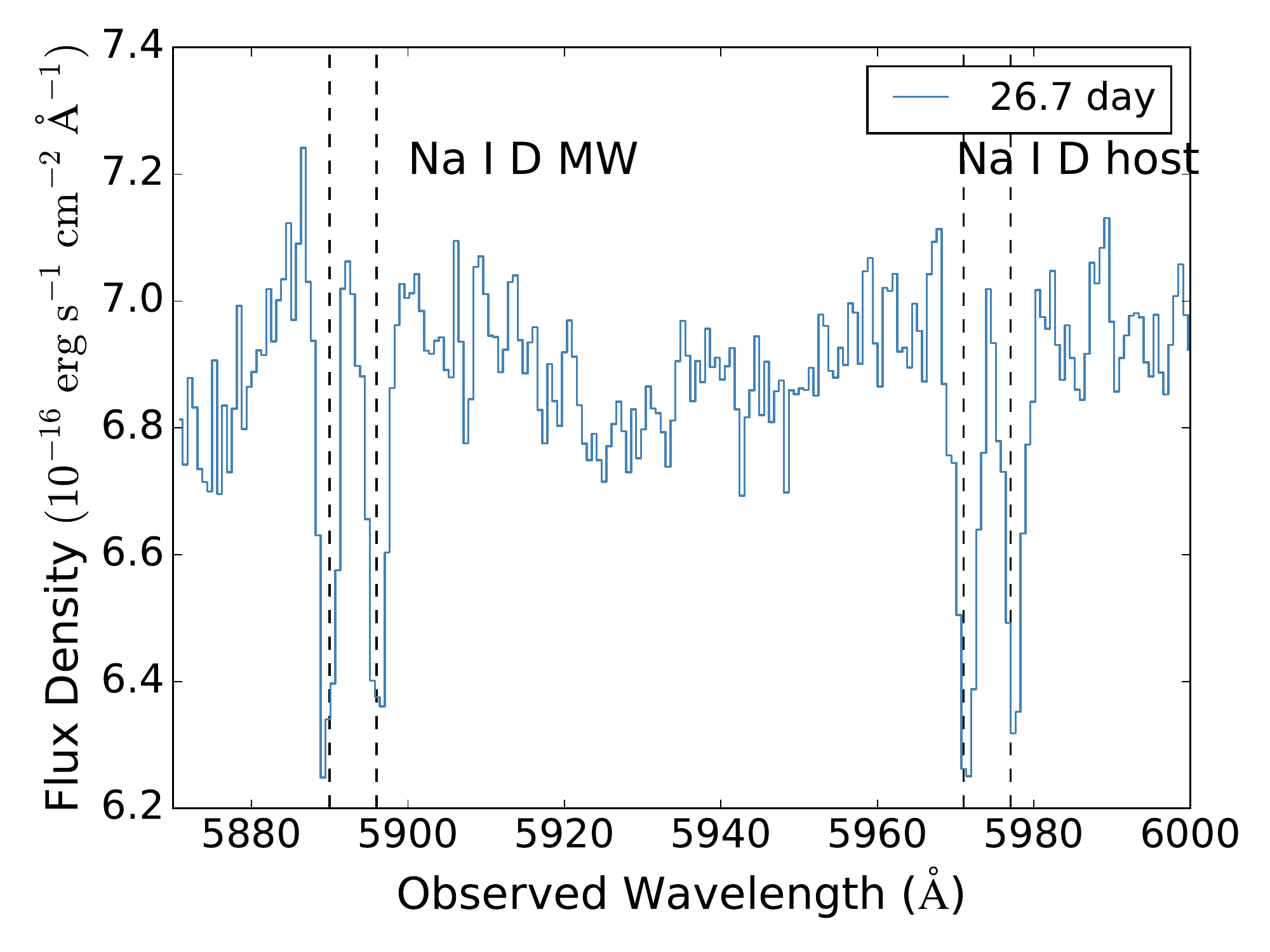}
\caption{Spectrum taken at $t \approx +26.7$\,d shows Na\,\uppercase\expandafter{\romannumeral1}\,D absorption from both the Milky Way (MW) and the host galaxy. \label{fig:extinction}}
\end{figure}

The Tully-Fisher distance of $60.1 \pm 1.4$\,Mpc \citep{2009ApJS..182..474S} from the NASA Extragalactic Database (NED) and a redshift $z = 0.013654$ estimated from H~{\sc ii} region emission lines in the host galaxy are adopted throughout this work. 
The Milky Way reddening is found to be $ E(B-V)_{\rm MW} = 0.059$\,mag for SN~2015bf \citep{2011ApJ...737..103S}, corresponding to a $V$-band extinction of 0.183\,mag assuming the \cite{1989ApJ...345..245C} extinction law ($R_V = 3.1$).
The equivalent width (EW) of the host-galaxy Na\,\uppercase\expandafter{\romannumeral1}\,D absorption feature is 0.62\,\AA\ on day 26.7 (see Fig. \ref{fig:extinction}), suggesting ${E(B-V)_{\rm host}} = 0.089$\,mag using the empirical relation derived by \citet{2003fthp.conf..200T}. 
Therefore, the total reddening of SN 2015bf is estimated to be ${E(B-V)_{\rm total}} = 0.148$\,mag. 

\subsection{Photometry} \label{subsec:photometry}
After the discovery of SN 2015bf, follow-up optical photometric observations started quickly using several telescopes, including the 80\,cm Tsinghua-NAOC Telescope (TNT) at Xinglong Observatory in China \citep{2012RAA....12.1585H}, the 2.4\,m Lijiang Telescope(LJT) of Yunnan Astronomical Observatory (YNAO) in China \citep{2015RAA....15..918F}, the 76\,cm Katzman Automatic Imaging Telescope (KAIT) located at Lick Observatory \citep{1993PASP..105.1164R,2001ASPC..246..121F}, and the Low Resolution Imaging Spectrometer (LRIS; \citealp{1995PASP..107..375O}) on the 10\,m Keck-I telescope in Hawaii.
These data covered the phases from $\sim 2$\,d to $\sim 205$\,d after explosion, and final flux-calibrated magnitudes are presented in Tables \ref{tab:clear} and \ref{tab:photometry1}.

All CCD images were pre-processed following standard routines, including bias subtraction, flat-field correction, and cosmic-ray removal \citep{2010ApJS..190..418G,2019MNRAS.490.3882S}. 
The template-subtraction technique was applied to the photometry.  
We performed point-spread-function (PSF) photometry for both the SN and the reference stars using DAOPHOT \citep{1987PASP...99..191S} or the pipeline {\tt Zrutyphot} (Mo et al. in prep.) developed for automatic photometry of images collected by TNT and LJT.
As there was no template for the $R$-band image obtained with the Keck-I telescope, 
we multiplied the total flux in the PSF by a scale factor, treated this as the flux of the SN, and then subtracted this from the total flux. We examined the excess flux to see if its distribution is consistent with the pre-SN image.
We tried different scale factors and finally adopted the value 0.45 (see Fig.\ref{fig:keck_image}), which gave a measurement of $21.691 \pm 0.128$\,mag. The quoted error are obtained by using scale factors of 0.4 and 0.5, respectively, which can be used as the systematic errors associated with the above method. 
To test the reliability of such an analysis, we further applied the image subtraction technique by using the pre-SN image from PS1, though it was obtained in r'-band. The resultant value is $21.844 \pm 0.086$\,mag, which is consistent with the above result. 

The instrumental magnitudes of the SN were converted into standard Johnson $UBV$ \citep{1966CoLPL...4...99J} and Kron-Cousins $RI$ \citep{1981SAAOC...6....4C} using a series of Landolt \citep{1992AJ....104..372L} and PS1 \citep{2016arXiv161205560C,2016arXiv161205243F,2016arXiv161205242M,2016arXiv161205245W} standard stars on a few photometric nights (see Table \ref{tab:standardstar}).
Apparent magnitudes obtained by KAIT were measured in the KAIT4 natural system.
The final results were transformed to the standard system using local
calibrators and colour terms for KAIT4 as given in Table 4 of \citet{2010ApJS..190..418G},
except for the KAIT clear-band data where no reliable
colour term is measured owing to the broad response function. We therefore
presented the magnitude relative to the reference stars in Landolt $R$ magnitude \citep{2003PASP..115..844L}, which is similar to the KAIT clear band.



\begin{figure*}
\includegraphics[width=1.5\columnwidth]{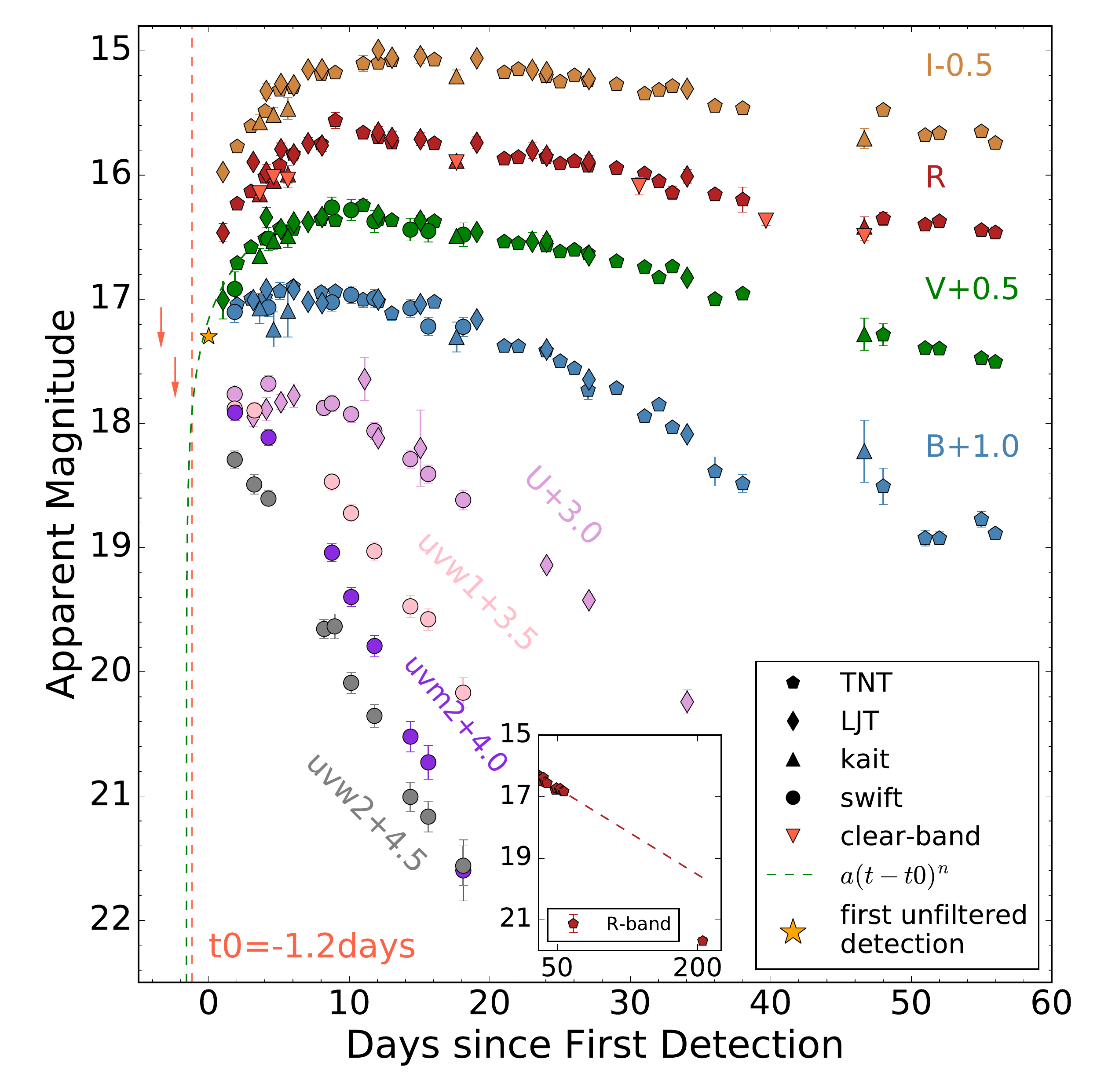}
\caption{The UV and optical light curves of SN 2015bf. The light curves of different bands have been vertically shifted for clarity. The vertical dashed line marks the explosion date estimate as the midpoint between the last nondetection and discovery. The green dashed line represents the $a(t-t_0)^n$ fits to the $V$-band magnitudes at early phases. The arrows represent the nondetection upper limits. The inset panel is the $R$-band light curve at late phases. The red dashed line in the inset is the linear fit for the $R$ data at $t \approx 50$\,d after explosion. \label{fig:lightcurve}}
\end{figure*}

\begin{figure*}

\subfigure{
\includegraphics[width=0.66\columnwidth]{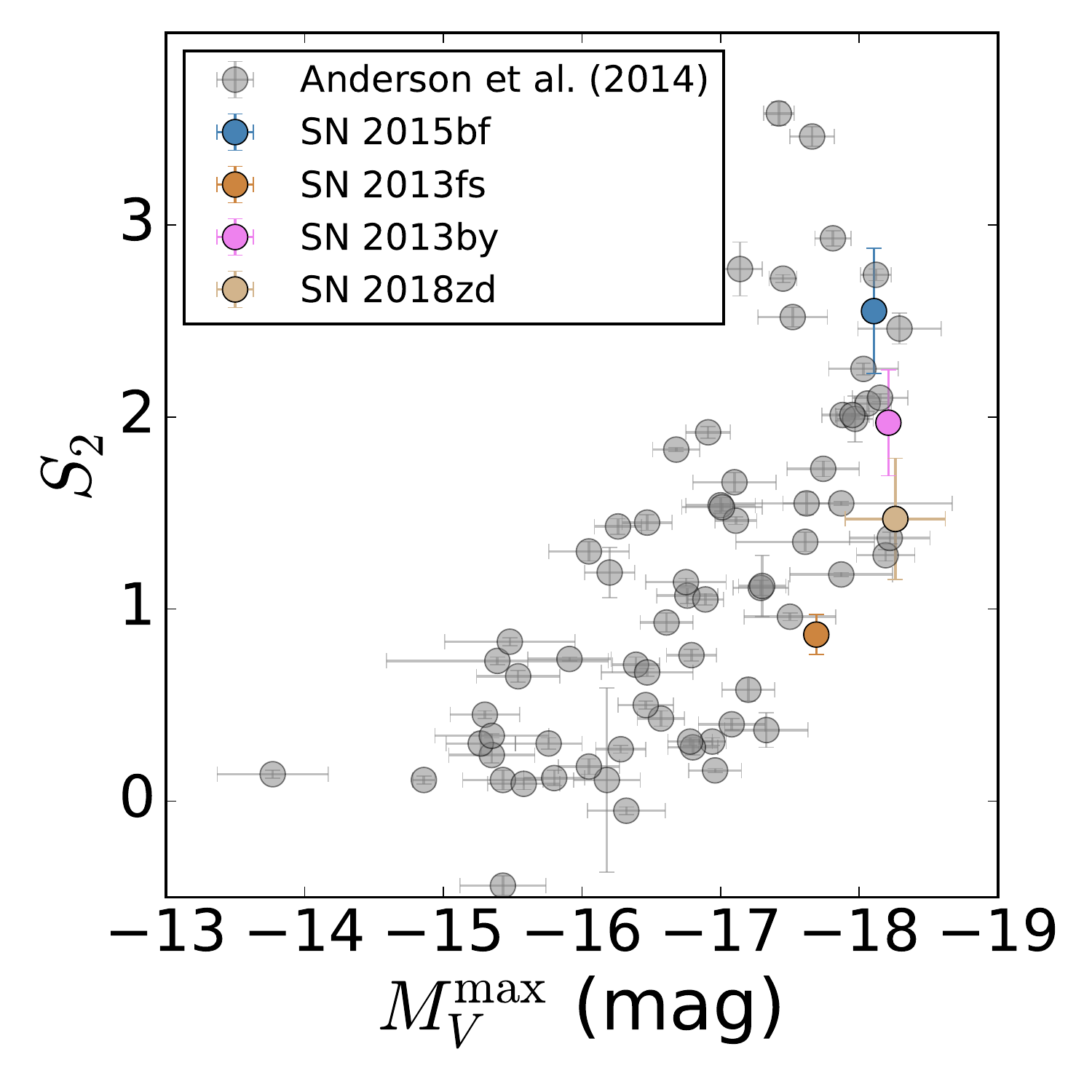}
}
\subfigure{
\includegraphics[width=0.66\columnwidth]{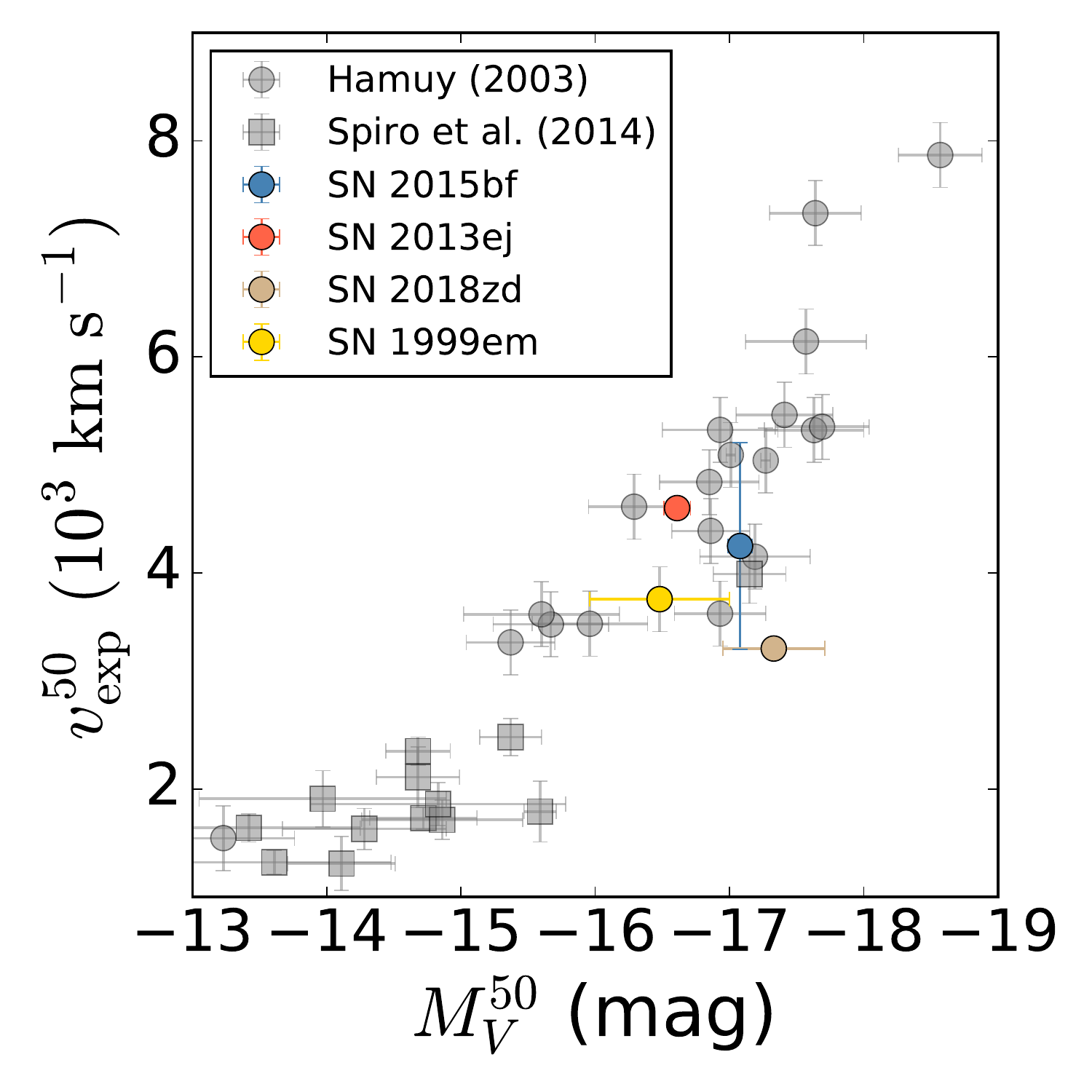}
}
\subfigure{
\includegraphics[width=0.66\columnwidth]{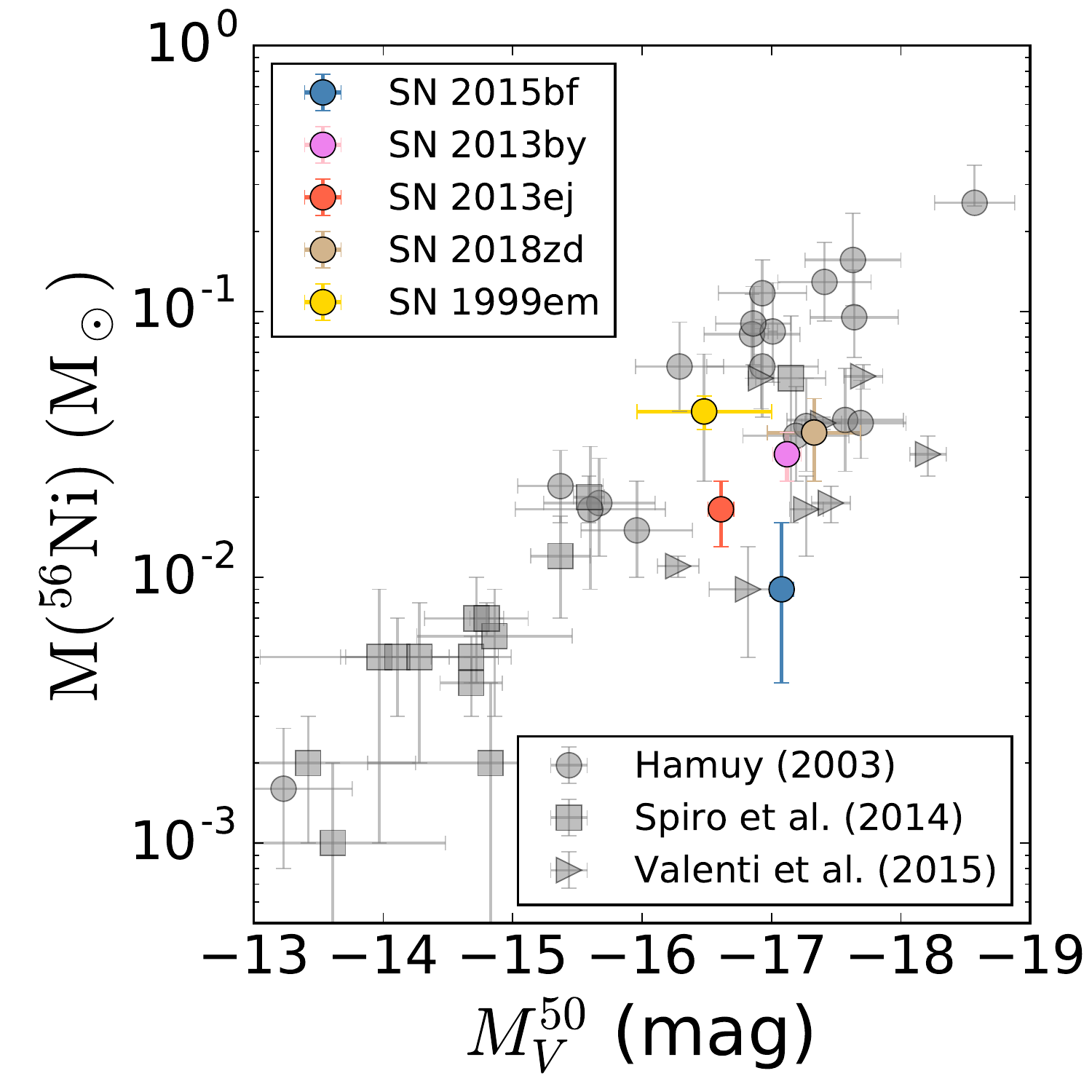}
}


\caption{The position of SN 2015bf in the SN~II family in terms of various photometric and spectroscopic indicators, including the maximun $V$-band absolute magnitude ($M^{\rm max}_V$), the decline rate of the plateau ($s_2$), 
the velocity of Sc~{\sc ii} $\lambda$6246 or Fe~{\sc ii} $\lambda$5169 measured 50\,d after explosion ($v^{\rm 50}_{\rm exp}$), 
the $V$-band absolute magnitude measured 50\,d after explosion ($M^{\rm 50}_V$), and the $^{56}{\rm Ni}$ mass. The grey dots represent the sample collected from \citet{2014ApJ...786...67A}, \citet{2003ApJ...582..905H}, \citet{2014MNRAS.439.2873S}, and \citet{2015MNRAS.448.2608V}.  SN 1999em, SN 2013ej, SN 2013by, and SN 2018zd are shown for comparision.\label{fig:photspeccompare}}
\end{figure*}

\begin{figure*}
\centering
\includegraphics[width=1.5\columnwidth]{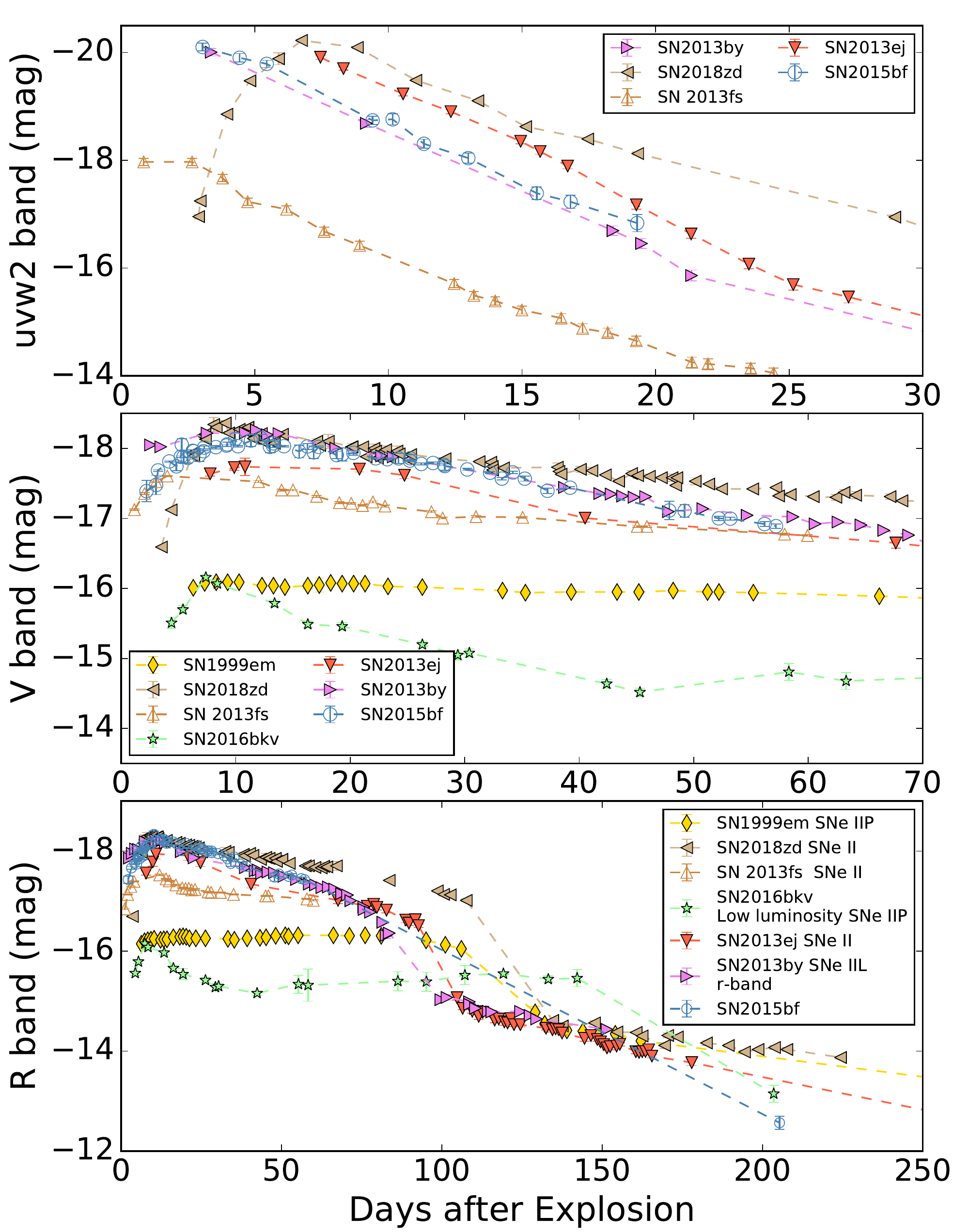}
\caption{Light-curve comparisons (in absolute magnitude) between SN 2015bf and some well-studied SNe~II, including the standard SN~IIP 
1999em \citep{2001ApJ...558..615H,2002PASP..114...35L,2003MNRAS.338..939E}, the fast-declining SNe~II 2013ej \citep{2015ApJ...807...59H,2016ApJ...822....6D,2016MNRAS.461.2003Y} and 2013by \citep{2015MNRAS.448.2608V}, and SNe~II showing interaction signatures in their spectra such as 
 SN 2013fs \citep{{2017NatPh..13..510Y},{2018MNRAS.476.1497B}}, 
SN 2016bkv \citep{{2018ApJ...859...78N},{2018ApJ...861...63H}}, and SN 2018zd \citep{2020MNRAS.498...84Z}.
\label{fig:comparelightcurve}}
\end{figure*}

\begin{figure}
\centering
\includegraphics[width=0.9\columnwidth]{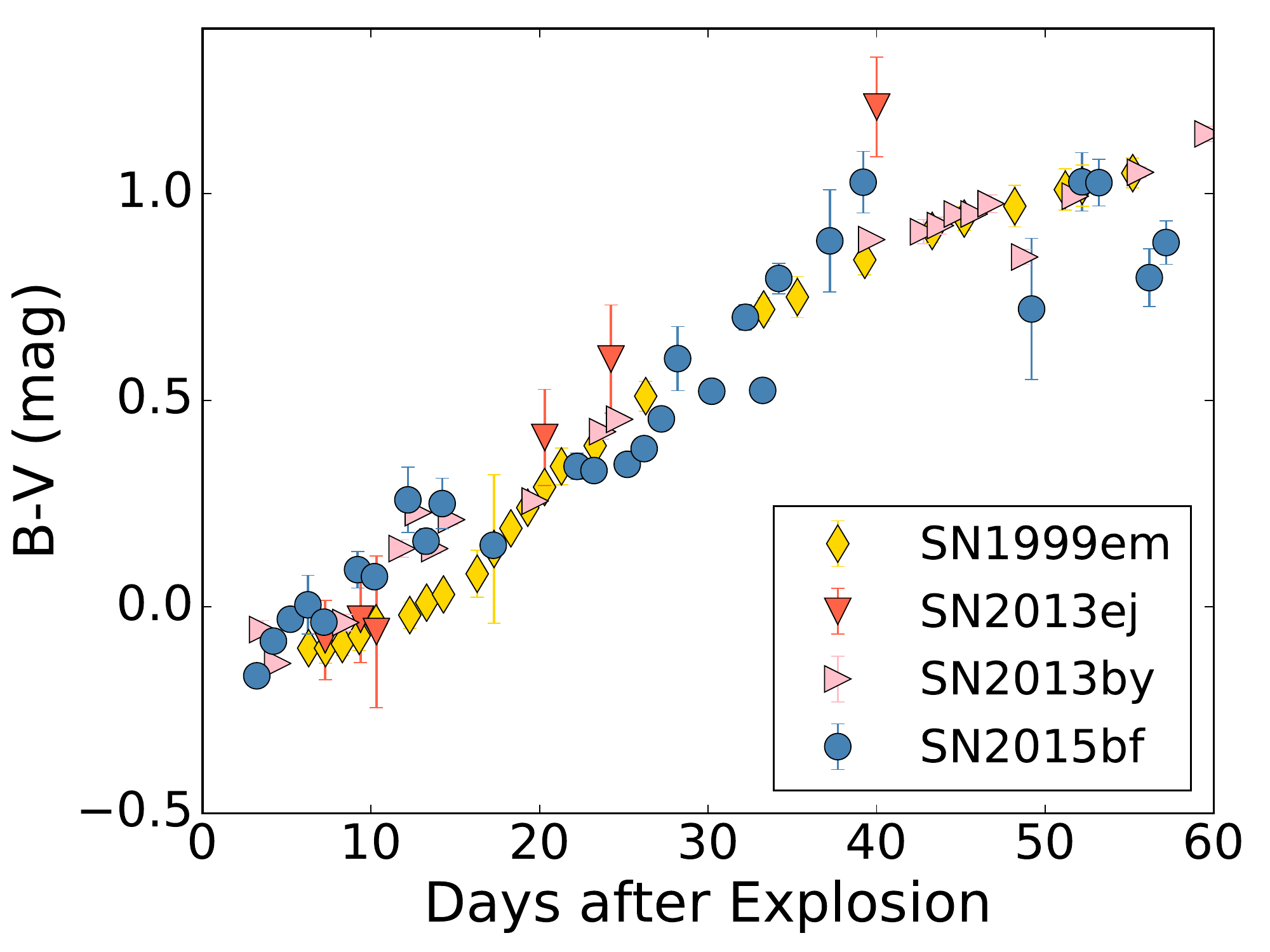}
\caption{ $B-V$ colour evolution of SN 2015bf , compared with that of SN 2013by, SN 2013ej, and SN 1999em.  \label{fig:colorcurve}}
\end{figure}

SN 2015bf was also observed in the $uvw2$, $uvm2$, $uvw1$, $u$, $b$, and $v$ bands using the Ultraviolet/Optical Telescope (UVOT) onboard the {\it Neil Gehrels Swift Observatory} \citep{2004ApJ...611.1005G,2005SSRv..120...95R}. Data reduction followed that of the {\it Swift} Optical Ultraviolet Supernova Archive \citep{2009AJ....137.4517B,2014Ap&SS.354...89B,2011AIPC.1358..373B}. 
A $3^{\prime\prime}$ or $5^{\prime\prime}$ aperture was used for the photometry after subtracting  the galaxy counts from a template image. The resulting UVOT photometry is listed in Table \ref{tab:swift}. 

\subsection{Spectroscopy} \label{subsec:spectroscopy}
15 optical spectra of SN 2015bf were collected using the Xinglong 2.16\,m telescope (+BFOSC), the LJT telescope (+YFOSC), the Lick 3\,m Shane telescope (+Kast; \citealp{Miller1993}), the Keck-II 10\,m telescope (+DEIMOS; \citealp{2003SPIE.4841.1657F}), and the Keck-I 10\,m telescope (+LRIS; \citealp{1995PASP..107..375O}). 
They mostly covered phases from $\sim 2$ to $\sim 50$\,d after the explosion, though a nebular spectrum was obtained at $t = 205$\,d. Details of the spectroscopic observations are listed in Table \ref{tab:spectra}. 

All spectra were reduced using standard IRAF pipelines. The procedures included bias and flat-field corrections, cosmic-ray removal, wavelength calibration, and flux calibration. 
The spectra were corrected for atmospheric extinction using the extinction curves of the local observatories. Telluric absorption was removed through comparison with the standard-star spectra.


\section{Photometric Properties} \label{sec:phot}

\subsection{Light Curve and Colour Curve} \label{subsec:lccc}

Figure \ref{fig:lightcurve} shows the ultraviolet (UV) and optical light curves of SN 2015bf.
The last nondetection in archival images was obtained with KAIT about 2.4\,d before the first detection. 
We thus adopt the midpoint ($t_0 = 2,457,367.77^{+1.2}_{-2.4}$) between the last nondetection  and the discovery dates as the estimated explosion time of SN 2015bf, though the upper limit of 17.83\,mag in the clear band does not place a tight constraint (the lower limit of explosion time is estimated by a simple power-law \citep{2015MNRAS.451.2212G} fit to the first three data points in $V$ and $R$ bands, see Figure \ref{fig:powerlaw_fit}, and the upper limit of explosion time is the discovery date).


SN 2015bf reaches the $V$-band peak with a rise time of $\sim 11$\,d, and the peak is $M_V = -18.11 \pm 0.08$\,mag. The $R$-band peak, $M_R = -18.23 \pm 0.10$\,mag, is also consistent with the trend that SNe~II with flash-ionised features tend to be luminous \citep{2016ApJ...818....3K}. 
After peaking, the $V$ light curve drops by $1.22 \pm 0.09$\,mag within the following $\sim 50$\,d.

In order to further describe the photometric properties of SN 2015bf, we compared the $V$-band absolute peak magnitude and $s_2$ parameter (magnitude decline rate per 100\,d measured for the second slope of the light curve) of SN 2015bf with those of the SN~II sample of \cite{2014ApJ...786...67A}. As shown in Figure \ref{fig:photspeccompare}, SN 2015bf belongs to the brighter and faster declining side of SNe~II. 
\cite{2014MNRAS.445..554F} proposed that SNe~IIL seem to decline by more than 0.5\,mag from peak brightness by day 50. Based on such a criterion, SN 2015bf should be classified as a SN~IIL because of its relatively quick decline ($1.03 \pm 0.12$\,mag after peak). 
Moreover, we noticed that the $R$-band magnitude at $t \approx 205$\,d is fainter than that expected from a linear decline by $2.1 \pm 0.13$\,mag.
In Figure \ref{fig:comparelightcurve}, we compare the $uvw2$, $V$, and $R$ light curves of SN 2015bf with those of other well-studied SNe~II, 
One can see that the light-curve properties (i.e., peak brightness and decline rate after maximum) of SN 2015bf are very similar to those of the fast declining SN 2013by and SN 2013ej.
If SN 2015bf has the same decline rate during the radioactive tail as SN 2013ej, then the $R$-band magnitude of SN 2015bf should drop $3.1 \pm 0.13 $\,mag
from $t \approx 57$\,d to $t \approx 100$\,d. 
\cite{2015MNRAS.448.2608V} suggested that all SNe~IIL exhibit a similar drop in the light curve as SNe~IIP, supporting the idea that SNe~IIP and SNe~IIL share similar underlying physics. 
Thus, it is possible that SN 2015bf had experienced a significant flux drop before entering the radioactive tail. 
Nevertheless,\cite{2015MNRAS.448.2608V} also showed that some SNe II which decline as fast as SNe IIL do not show significant flux drop, though such sample is rare. This indicates an overall gradual flux decline after the peak cannot be ruled out for SN 2015bf due to the lack of abservations between $t \approx 57$ days and $t \approx 205$ days.



The $B-V$ colour curve of SN 2015bf is shown in Figure \ref{fig:colorcurve}, together with that of SN 1999em, SN2013ej, and SN 2013by. 
The evolution of its colour curve is similar to that of other SNe~II.
At early phases, the colour of SN 2015bf is blue. At $\sim 10$\,d, the $B-V$ colour starts to evolve toward the red as the photosphere expands and cools.

\subsection{Spectral Energy Distribution} \label{subsec:sed}
\begin{figure*}
\centering
\includegraphics[width=1.5\columnwidth]{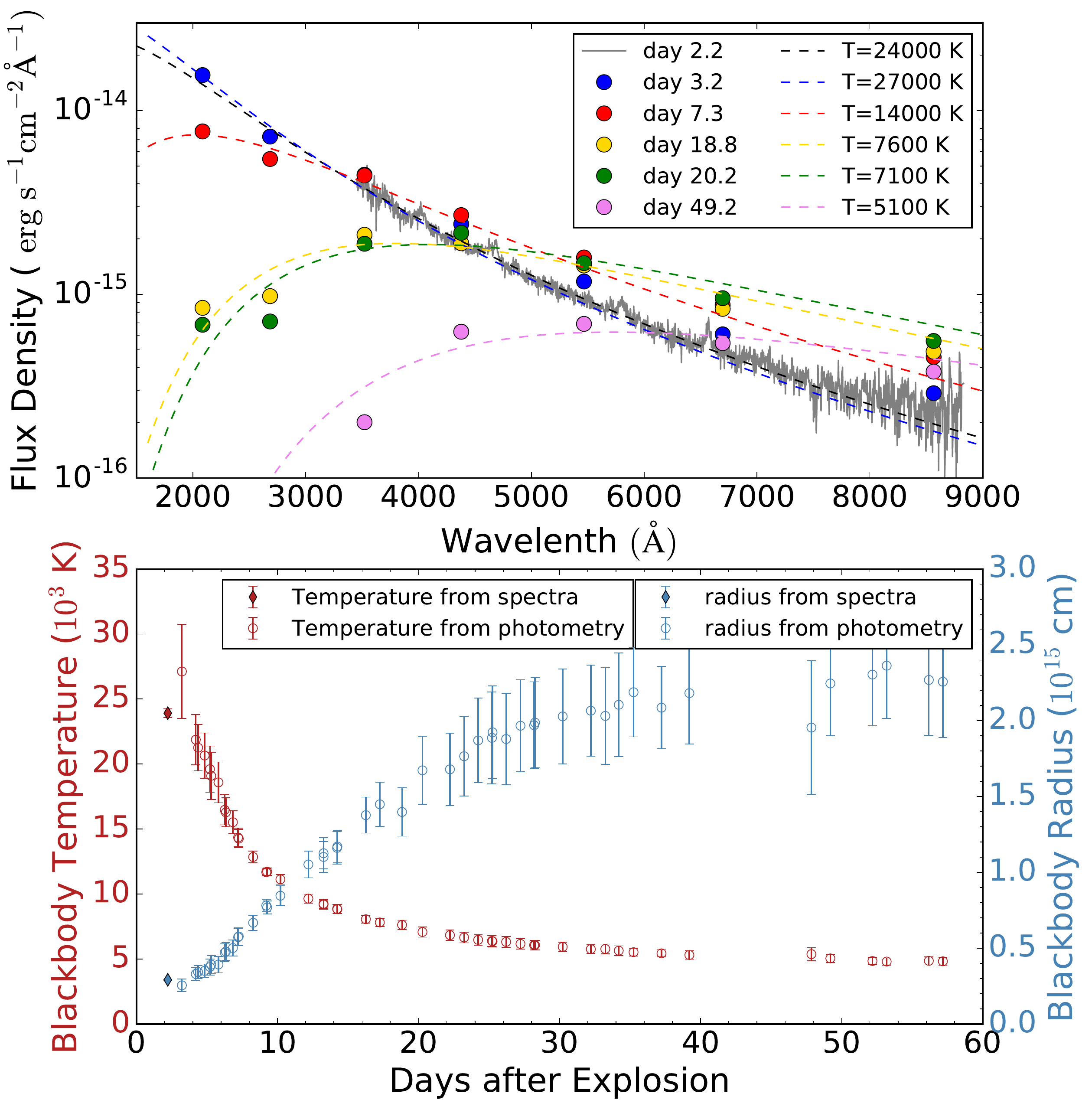}
\caption{{\it Upper panel:} SED of SN 2015bf at selected phases. The dashed lines are blackbody fits to the observations. {\it Lower panel:} Evolution of the balckbody temperature and radius of SN 2015bf. \label{fig:sed}}
\end{figure*}

Using the UV and optical photometry, we constructed the spectral energy distribution (SED) and derived the temperatures and radii by applying blackbody fits. Since the phases of {\it Swift} observations do not match well with those of the $BVRI$ observations, we interpolated $uvw2$, $uvw1$, and $u$ data to the epochs when the optical observations are available. However, the {\it Swift} observations only last for 19 days; after that, we extrapolated the {\it Swift} observations linearly. 
We also fit a blackbody to the spectrum at $t = 2.2$\,d as it was obtained earlier than the photometry. 
The SEDs and their blackbody fits are shown in the upper panel of Figure \ref{fig:sed}, while the evolution of blackbody temperatures and radii is illustrated in the lower panel.

\citet{2018ApJ...861...63H} noticed a possible temperature rise during the first six days for SN 2016bkv which might be related to the flash ionization of CSM. 
Such temperature increase was also observed in SN 2018zd \citep{2020MNRAS.498...84Z}.
For SN 2015bf, we have only one data point before the peak value, which does not allow us to judge its temperature evolution at early phases due to the lack of the UV constraints.

\section{Optical Spectra} \label{sec:spec}
\subsection{Evolution of Spectra}\label{subsec:evolution_of_spectra}

\begin{figure}
\includegraphics[width=\columnwidth]{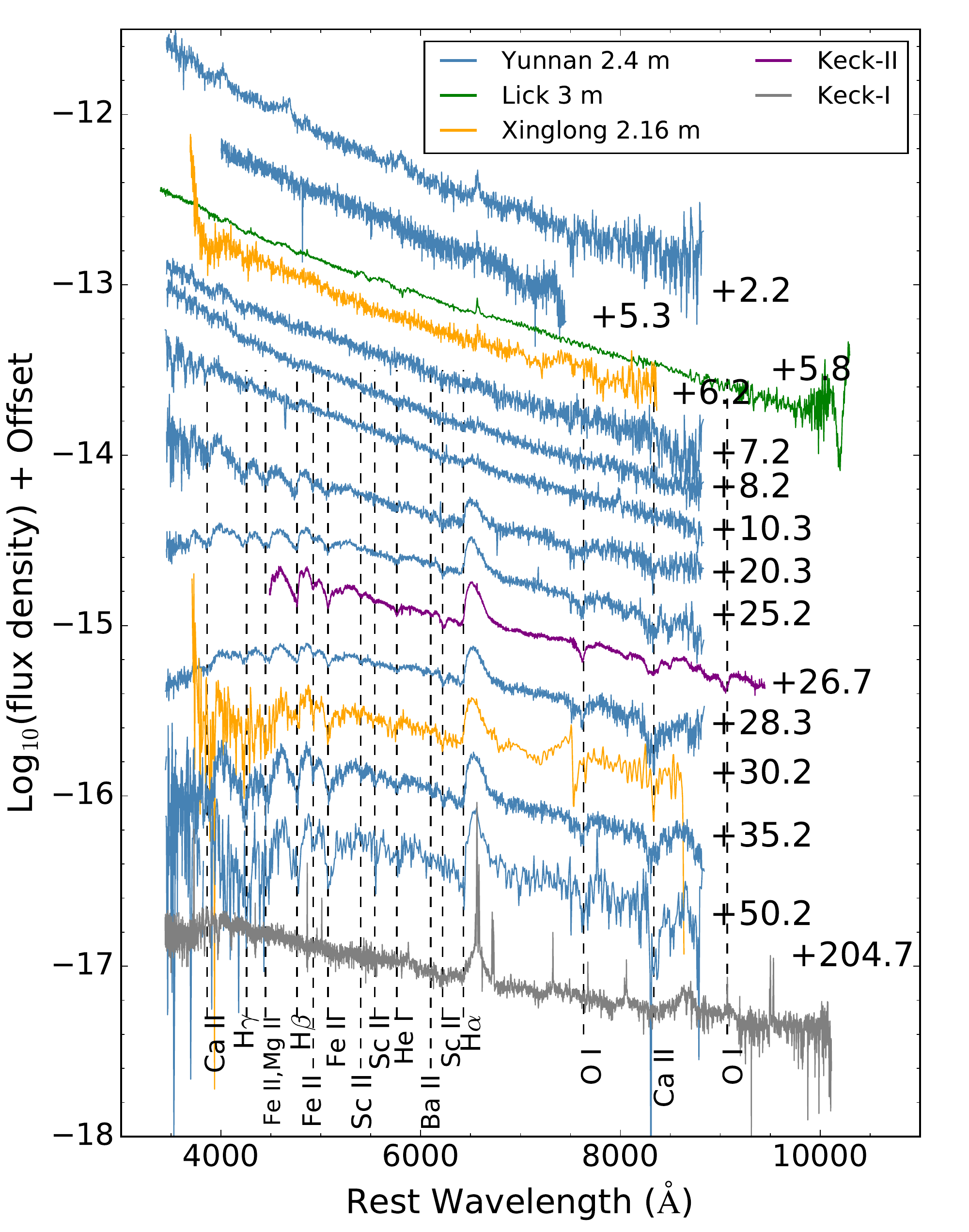}
\caption{Optical spectra of SN 2015bf, shifted vertically for clarity. All of the spectra have been corrected for host-galaxy and Milky Way extinction. The numbers on the right side mark the epochs of the spectra in days after the explosion. The continuum of the spectra on days +5.3, +20.3, +25.2, and +28.3 has been adjusted with the broadband $UBVRI$ photometry at similar phases. \label{fig:spectra}}
\end{figure}

\begin{figure}
\includegraphics[width=\columnwidth]{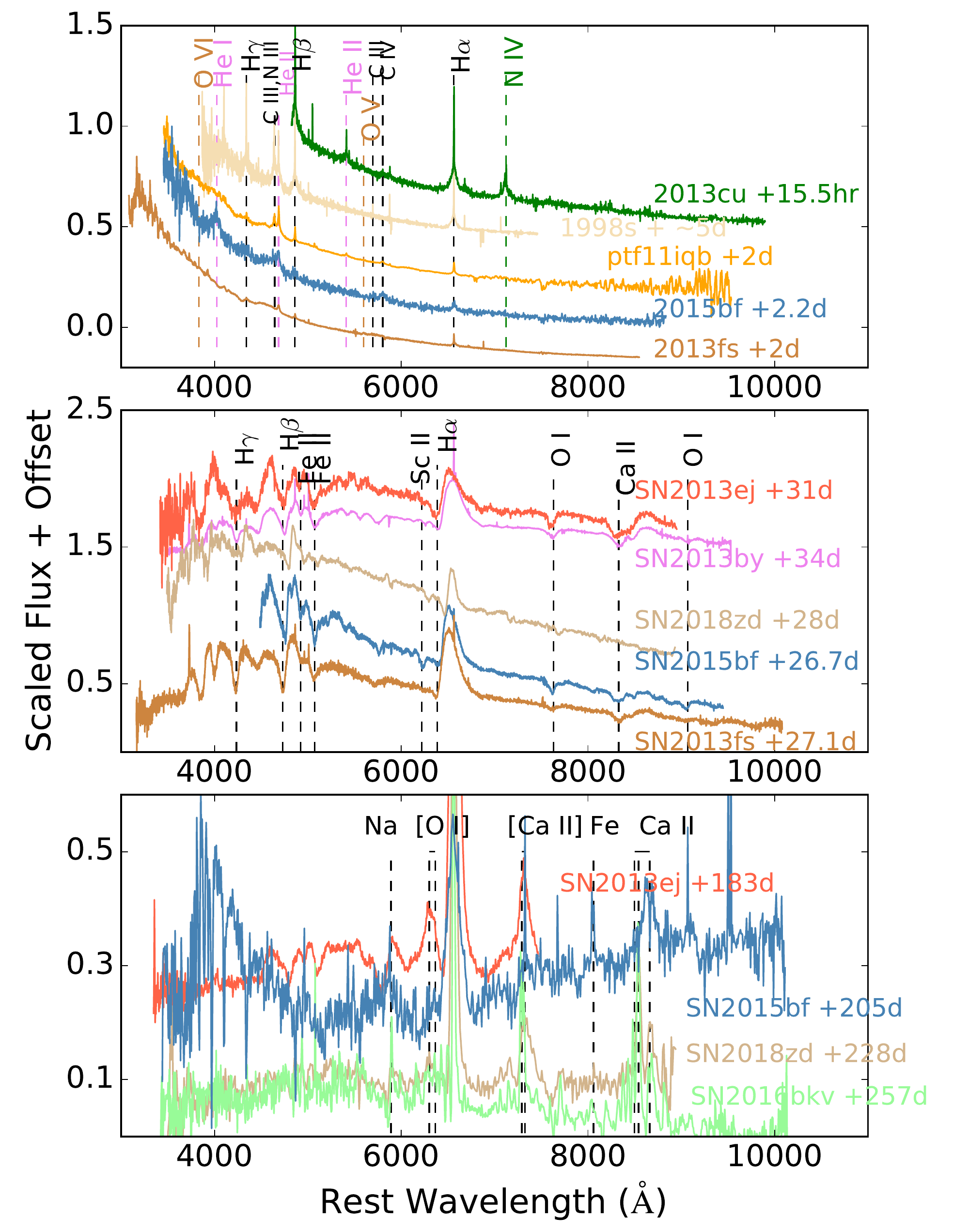}
\caption{Spectral comparisons of SN 2015bf with SN 1998S \citep{2015ApJ...806..213S}, 
PTF11iqb \citep{2015MNRAS.449.1876S}, SN 2013cu \citep{2014Natur.509..471G}, SN2013fs \citep{2017NatPh..13..510Y}, SN 2013ej \citep{2015ApJ...807...59H,2016MNRAS.461.2003Y}, SN 2013by \citep{2015MNRAS.448.2608V}, SN 2016bkv \citep{2018ApJ...861...63H}, and SN 2018zd \citep{2020MNRAS.498...84Z} at early, photospheric, and nebular phases (top to bottom panels). A blackbody continuum ($T = 7500$\,K) has been subtracted from the nebular spectrum of SN 2015bf; narrow emission lines of H$\alpha$, [N~\uppercase\expandafter{\romannumeral2}] $\lambda\lambda$6548, 6583, [O~\uppercase\expandafter{\romannumeral2}] $\lambda3727$, and [O~\uppercase\expandafter{\romannumeral3}] $\lambda\lambda$4959, 5007 from the host galaxy were also removed.  \label{fig:spectracompare}}
\end{figure}

The 15 optical spectra of SN 2015bf span from $t = 2.2$ to 205\,d after the estimated explosion date. The complete spectral evolution is shown in Figure \ref{fig:spectra}. Figure \ref{fig:spectracompare} shows the spectral comparison of SN 2015bf with some representative SNe~II at early, photospheric, and nebular phases. The comparison sample includes SN 1998S \citep{2015ApJ...806..213S}, PTF11iqb \citep{2015MNRAS.449.1876S}, SN 2013cu \citep{2014Natur.509..471G}, SN 2013fs \citep{2017NatPh..13..510Y}, SN 2013ej \citep{2015ApJ...807...59H,2016MNRAS.461.2003Y}, SN 2013by \citep{2015MNRAS.448.2608V}, SN 2016bkv
\citep{2018ApJ...861...63H}, and SN 2018zd \citep{2020MNRAS.498...84Z}. 

\begin{figure}
\centering
\includegraphics[width=\columnwidth]{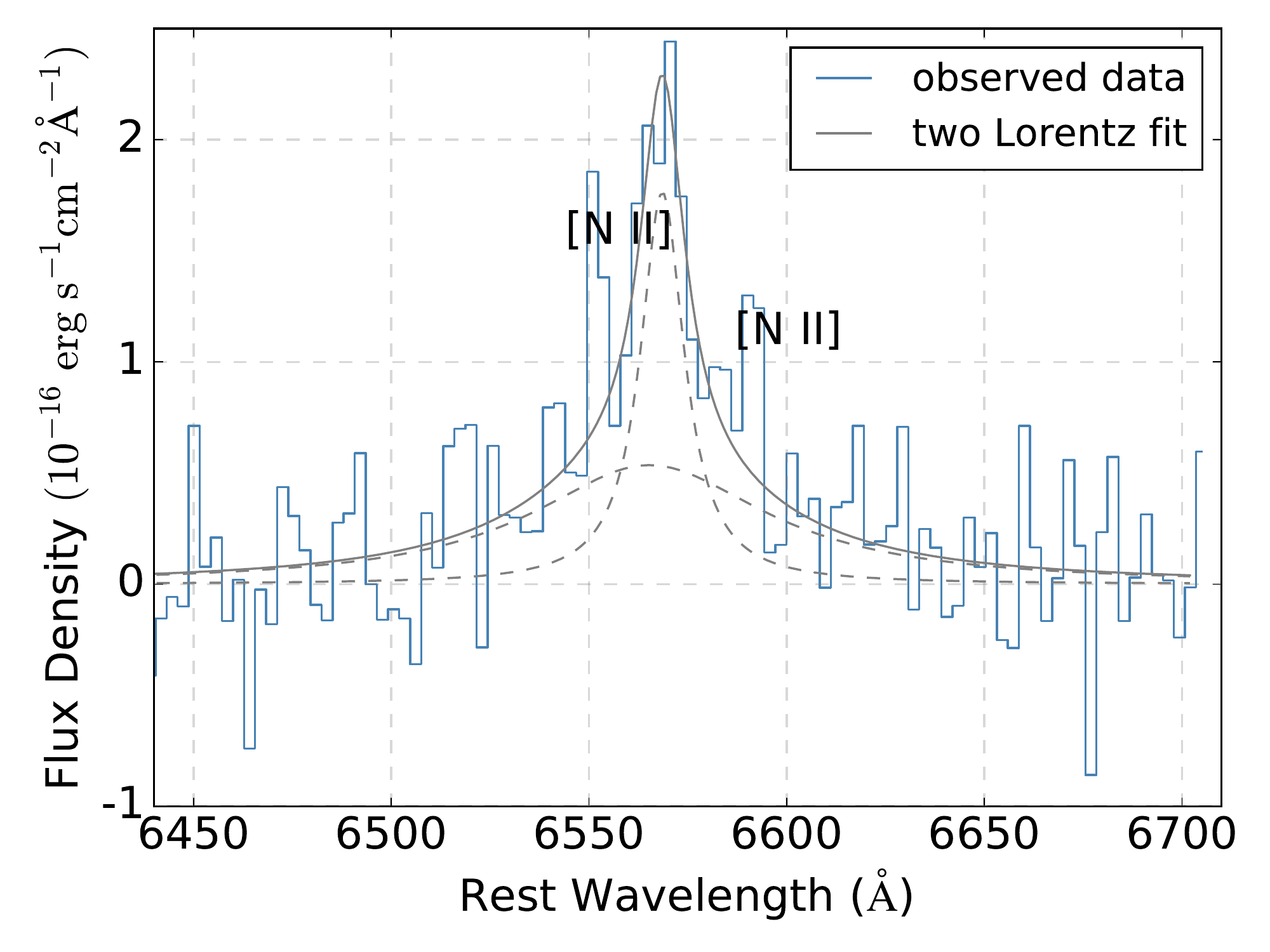}
\caption{The H$\alpha$ line profile in the spectrum of SN 2015bf, taken at $t \approx 2.2$\,d. The grey dashed lines are the Lorentzian fits for the narrow and broad components of the H$\alpha$ line profile. The narrow emission features on both sides of H$\alpha$ are [N~{\sc ii}] lines from the host galaxy.   \label{fig:ha}}
\end{figure}

The first spectrum is characterised by a very blue continuum with prominent emission lines, including H$\alpha$, H$\beta$, 
He~{\sc i} $\lambda$4026, C~{\sc iv} $\lambda$5801, and blends of He~{\sc ii} $\lambda$4686 and 
C~{\sc iii}/N~{\sc iii}. 
The H$\alpha$ feature shows a unique line profile with a narrow core and broad Lorentzian wings, as shown in Figure \ref{fig:ha}. Such FI emission lines are formed from the recombination of CSM ionised by high-energy photons created by SN shock breakout  \citep{2014Natur.509..471G,2016ApJ...818....3K}. They have been seen in a few other SNe at early times, and their comparisons with SN 2015bf are shown in the upper panel of Figure \ref{fig:spectracompare}. 

The FI features usually exist in very early-time spectra, and they do not last for a long period because the nearby CSM responsible for them is destroyed by the expanding ejecta. In SN 2015bf, we cannot see any emission lines 7.2\,d after explosion, except for the faint H$\alpha$ which is likely contributed by the H~{\sc ii} region. By $t \approx 10$\,d, the spectra evolve like those of a regular SN~II with broad P-Cgyni profiles. 
As the ejecta expand and the photosphere recedes, we identify absorption features of Sc~{\sc ii}, Mg~{\sc ii}, Fe~{\sc ii}, Ba~{\sc ii}, Ca~{\sc ii}, and O~{\sc i}; see Figure \ref{fig:spectra}. 
Comparing with SN 2013ej, SN 2013by, SN 2013fs, and SN 2018zd at the photospheric phase, one can see that these SNe share similar spectral features. 
Nevertheless, the H$\alpha$ line profile of SN 2015bf has a weaker absorption component in comparision with that of SN 2018zd and SN 2013fs. 
Previous studies showed that there is a tendency for brighter SNe~II with faster post-peak declines to have smaller ratios of absorption to emission in H$\alpha$ \citep{1994A&A...282..731P,1996AJ....111.1660S}. This can be explained as a result of a less-massive H envelope \citep{2014MNRAS.445..554F,2014ApJ...786L..15G} or as an effect of CSM interaction.


Intermediate-width forbidden emission lines characterise nebular-phase spectra of SNe~II. For SN 2015bf, we could identify features of [O~{\sc i}], H$\alpha$, Ca~{\sc ii}, Fe~{\sc i}, and  Na~{\sc i}. Compared with SN 2013ej and SN 2018zd at a similar phase, the nebular spectrum of SN 2015bf at $t \approx 205$\,d is relatively featureless, with less evident emission lines 
of [O~{\sc i}] $\lambda\lambda$6300, 6364, Ca~{\sc ii} $\lambda$7291, and Ca~{\sc ii} $\lambda\lambda$8498, 8452. 
As the main indicator of the progenitor mass is the [O~{\sc i}] $\lambda\lambda$6300, 6364 doublet \citep{2012A&A...546A..28J}, the weak [O~{\sc i}] features of SN 2015bf indicate that its progenitor mass could differ from that of SN 2013ej, although their light-curve properties are similar. 


\subsection{Ejecta Velocity}\label{subsec:velocity}
\begin{figure}
\centering
\includegraphics[width=\columnwidth]{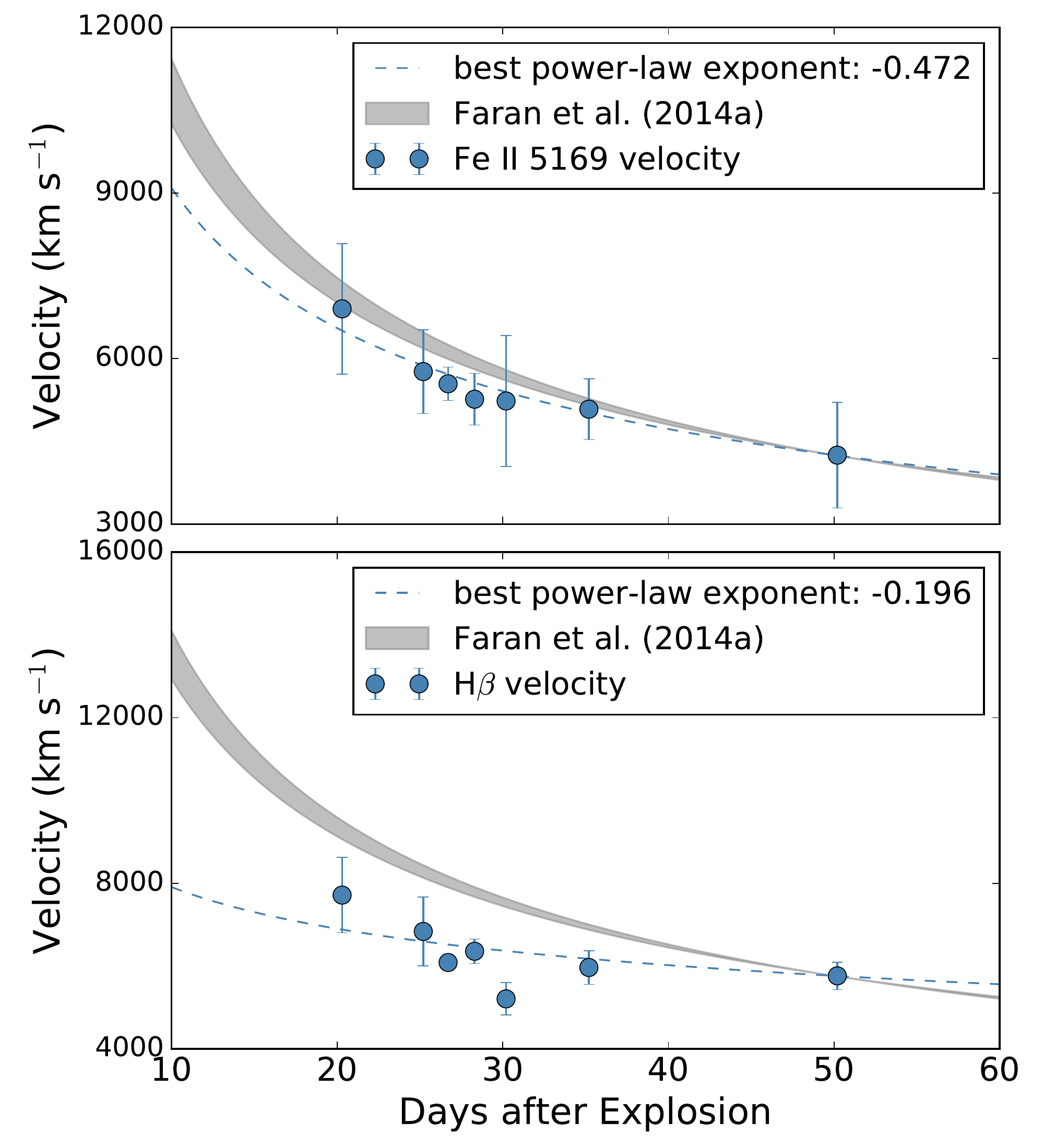}
\caption{The velocity evolution of H$\beta$ and 
Fe~{\sc ii} $\lambda$5169. 
The blue dashed lines represent the power-law fit for the H$\beta$ and Fe~{\sc ii} velocities. 
The grey shaded area represents the power-law fit for 
SNe~IIP from \citet{2014MNRAS.442..844F}. \label{fig:velocity}}
\end{figure}

For the spectra covering the phases from day 20 to 50 after explosion, the ejecta velocity can be inferred from the absorption minima of the P~Cygni profiles. 
As the absorption feature of H$\alpha$
is not noticeable, we only measured the velocity from the H$\beta$ and Fe~{\sc ii} $\lambda$5169 lines. The velocity evolution with the best power-law fit is shown in Figure \ref{fig:velocity}. For comparision, we also indicate the power-law fit $v_{\rm {Fe}} = v_{50}(t/{50})^{-0.581 \pm 0.034}$ and 
$v_{\rm H\beta}=v_{50}(t/{50})^{-0.529 \pm 0.027}$ from 23 SNe~IIP published by \citet{2014MNRAS.442..844F}. The best-fit power-law exponent for Fe~{\sc ii} is $-0.472 \pm 0.025$,
consistent with the exponent found by \cite{2014MNRAS.442..844F}. The photospheric velocity (indicated by Fe~{\sc ii} $\lambda$5169) measured at $t \approx 50$\,d after explosion and the $V$-band absolute
magnitude also follow the luminosity-velocity relation found for SNe~IIP  \citep{2003ApJ...582..905H}. However, the expansion velocity of H$\beta$ declines at a slower rate for SN 2015bf, with a best-fit index of $-0.196 \pm 0.06$, much smaller than the typical value found for SNe~IIP. Such slower velocity evolution for H$\beta$ is consistent with the result found for SNe~IIL by \citet{2014MNRAS.445..554F}; they
explained this as the H line being formed in the outer layers of the ejecta rather than in a gradually exposed H envelope. 


\section{Discussion} \label{sec:discussion}

\subsection{Presupernova Mass-Loss History}\label{subsec:flash spectra}
Direct observations of final-stage stellar evolution are very challenging. Flash-ionised features provide opportunities to speculate about the mass-loss history shortly before the SN explosion. For SN 2015bf, considering the ejecta velocity as $\sim 10,000$\,km\,s$^{-1}$, the disappearance of FI emission lines within the first week after explosion indicates that the CSM 
should be located at a distance $\lesssim 6 \times 10 ^{14}$\,cm.


The full width at half-maximum intensity (FWHM) of the H$\alpha$ narrow component in the first spectrum is $\sim 600$\,km\,s$^{-1}$. If we treat this value as the wind velocity, then the wind can reach the CSM location within 120\,d. That means the CSM started to form nearly 120\,d before explosion. 
Since the instrumental FWHM of the first spectrum (measured from the FWHM of the night-sky O~{\sc i} $\lambda$5577 emission line) is close to the FWHM of the H$\alpha$ narrow component, the wind velocity inferred from the (FWHM) of the H$\alpha$ narrow component will have  large uncertainty.
If we instead adopt the wind velocity to be $\sim 100$\,km\,s$^{-1} $ (as for SN 2013fs), then the CSM started to form $\sim 2$\,yr before the SN explosion.



Following the method described by \cite{2013ApJ...768...47O} and the relation derived by \cite{2006agna.book.....O}, we also estimate the mass-loss rate based on the luminosity of the narrow H$\alpha$ feature.  Assuming a temperature of $\sim 24,000$\,K (the blackbody temperature of the first spectrum of SN 2015bf), we can then obtain a relation 
\begin{small}
\begin{equation}
L_{\rm H\alpha} \leq (4.36 \times 10^{39}) {(\frac{\dot{M}}{\rm 10^{-2}\,M_\odot\,yr^{-1}})}^2 {(\frac{v_{\rm wind}}{\rm 500\,km\,s^{-1}})}^{-2} ({\frac{r_2-r_1}{r_1}}) {(\frac{r_1}{\rm 10^{15}\,cm})}^{-1}\,{\rm erg\,s^{-1}},
\end{equation}
\end{small}
where $r_1$ and $r_2$ represent the lower and upper limits of the H$\alpha$-emitting region, respectively. 
Here we adopt $r_1 \approx 2.9 \times 10^{14}$\,cm (the blackbody radius of the first spectrum of SN 2015bf) and $r_2 \approx 6 \times 10^{14}$\,cm.
According to the flux of the narrow component of H$\alpha$ ($3.6 \times 10^{-15}\,{\rm erg\,s^{-1}\,cm^{-2}}$) and adopting a distance of 60.1\,Mpc, we derive the H$\alpha$ luminosity as $L_{\rm H\alpha} = 1.56 \times 10^{39}\, {\rm erg\,s^{-1}}$. This luminosity suggests that the mass-loss rate should be larger than  $3.7 \times 10^{-3}\,{\rm M_\odot\,yr^{-1}}$. Such a value is much larger than the typical mass-loss rate of $10^{-6}$--$10^{-5}\,{\rm M_\odot\,yr^{-1}}$ for red supergiants (RSGs) \citep{2014ARA&A..52..487S}. Since the flux of H$\alpha$ is enhanced by the host-galaxy H~{\sc ii} region (see Fig. \ref{fig:ha}), the true luminosity of H$\alpha$ should be slightly lower,
$< 1.56 \times 10^{39}\,{\rm erg\,s^{-1}}$. However, the order of magnitude $10^{-3}\,{\rm M_\odot\,yr^{-1}}$ is consistent with the mass-loss rate of other SNe~II, such as SN 2013fs ($\sim 10^{-3}\,{\rm M_\odot\,yr^{-1}} $; \citealt{2017NatPh..13..510Y}). Therefore, we consider the derived mass-loss rate of SN 2015bf to be reliable, and we propose that the progenitor of SN 2015bf experienced violent mass loss shortly before exploding.

\cite{2016ApJ...818....3K} suggested that all of their FI events peak more luminous than $M_R = -17.6$\,mag and have SN~IIL-like initial decline after the peak. Moreover, they suggested that more luminous SNe tend to maintain higher temperatures during their early-time evolution. SN 2015bf is such an example favouring the above arguments. 
\cite{2017ApJ...838...28M} showed that the multiband light curves of SNe~IIL can be well fitted by ordinary RSGs surrounded by dense CSM produced a few months or years before the SN explosion.   
\citet{2019A&A...631A...8H} also proposed that increasing the mass of the CSM will lead to featureless spectra at early phases and weak H$\alpha$ absorption during the recombination phase, which is seen in SN 2015bf. 
Considering the high peak luminosity, the fast-declining light curve, the blue featureless spectra at early phases, the weak H$\alpha$ absorption in the photospheric phase, and a compact CSM envionment, interaction betweet ejecta and CSM may play an important role in modifying the observed light curves and spectra of SN 2015bf. 

\subsection{Bolometric Light Curve and Explosion Parameters}\label{subsec:explosion parameter}

\begin{figure}
\centering
\includegraphics[width=\columnwidth]{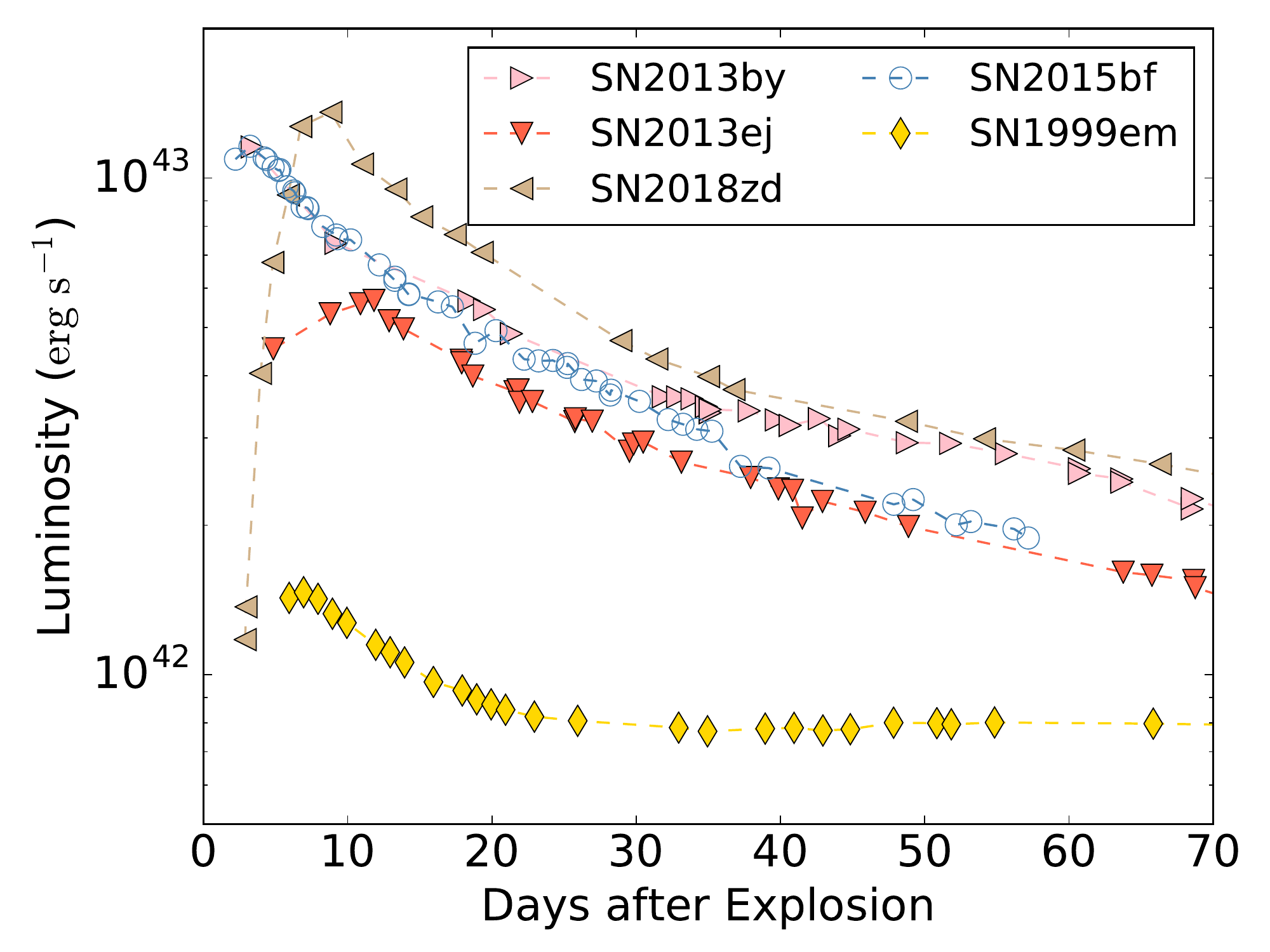}
\caption{Bolometric light curve of SN 2015bf, compared with those of other well-studied SNe~II. \label{fig:bolometric}}
\end{figure}

The bolometric luminosity of SN 2015bf was derived by integrating the SED from near-infrared through optical emission. Flux outside the wavelength range of photometric coverage was extrapolated based on a blackbody spectrum. Figure \ref{fig:bolometric} displays the bolometric light curve of SN 2015bf. The bolometric light curves of SN 1999em, SN 2013ej, SN 2013by, and SN 2018zd produced by the same method are also shown for comparison. 
Similar to SN 2013by and SN 2018zd, SN 2015bf has a relativly higher bolometric peak ($L_{\rm max} = 1.16 \times 10^{43}\,{\rm erg\,s^{-1}}$), which is $\sim  8$ times the peak luminosity of SN 1999em.

Assuming the progenitor of SN 2015bf is an RSG, we can estimate explosion parameters and properties of the progenitor based on empirical relations.   \cite{2016ApJ...829..109M} provide a relation between the $g$-band rise time and the radius at the time of explosion for SNe~II, as they found that the properties of the light curve at early phases depend mainly on the progenitor radius. Using their relation, we can calculate the progenitor radius of SN 2015bf. The $g$ magnitudes of SN 2015bf were converted from $B$ and $V$ data using the relation of \cite{2005AJ....130..873J}. The rise time in $g$ was estimated as $8.5^{+2.4}_{-1.2}$\,d, corresponding to a radius of  $677^{+241}_{-115}$\,R$_{\odot}$.

\cite{1985SvAL...11..145L} connected the mass of the ejecta, the progenitor radius, and the explosion energy to the plateau duration, the absolute $V$ magnitude, and the photosphere velocity at the mid-plateau epoch. Using their Equation (3), we derived the plateau duration as $56^{+10}_{-14}$\,d. With this plaueau duration, we derive the explosion energy to be $(0.35 \pm 0.17) \times 10^{51}\,{\rm erg\,s^{-1}}$ and the ejecta mass to be $3.05^{+2.15}_{-1.72}$\,M$_{\odot}$ using their Equations (1) and (2). 

\subsection{Faint Tail and Electron-Capture Supernova? }\label{subsec:ECSNe}


As seen in the lower panel of Figure \ref{fig:comparelightcurve}, SN 2015bf has a low luminosity in the nebular phase. The $R$ magnitude at $t \approx 205$\,d after explosion drops by $\sim 5$\,mag compared with the magnitude measured at $t \approx 50$\,d after explosion.
Although having a similar peak luminosity, the $R$ magnitude of SN 2015bf at $t \approx 205$\,d is $1.5 \pm 0.13$\,mag
fainter than that of SN 2018zd. Its tail luminosity is even fainter than that of SN 2016bkv, whose peak magnitude is much fainter than that of SN 2015bf. 

Based on the $R$-band magnitude, we estimated that the luminosity of SN 2015bf at $t \approx 205$\,d is only $0.25\pm{0.08}$ times that of SN 2018zd (the lower and upper limits are estimated when the scale factor is adopted as 0.3 and 0.6, respectively; see Fig. \ref{fig:keck_image}). Following the the second equation of \cite{2003ApJ...582..905H}, we estimated the nickel mass of SN 2015bf to be  $0.009^{+0.007}_{-0.005}\,{\rm M_\odot}$. (The lower and upper errors are $3 \sigma$ uncertainty).  
As seen in the right pinel of Fig. \ref{fig:photspeccompare}, SN 2015bf seems to have less nickel mass compared with other SNe II with similar $V$-band absolute magnitude measured 50 days after explosion. 
The equation (2) of \citet{2003ApJ...582..905H} assumes that all the gamma rays due to ${\rm ^{56}Co \to ^{56}Fe}$ are fully trapped. For most SNe IIP, this is a reasonable assumption as the decay rates on the tail are consistent with the value for full trapping. For SN 2015bf, we can not conclude whether all the gamma rays are fully trapped as we only have one data point at late times. If not, the nickel mass may be underestimated.


Electron-capture supernovae \citep{1982Natur.299..803N} are predicted to produce such a small amount of nickel mass ($<0.015 {\rm M_\odot}$) \citep{2006A&A...450..345K}.
Moreover, the progenitors of electron-capture SNe should explode within the CSM environment created by the super-asymptotic-giant-branch (AGB) wind \citep{2014A&A...569A..57M}.
SNe~IIn-P (SNe~IIn with plateau phase as seen in SNe~IIP \citep{2017hsn..book..403S}) have been suggested to be related to electron-capture SNe because of their low nickel mass \citep{1998ApJ...493..933S,2012MNRAS.424..855K,2013MNRAS.431.2599M}.
We noticed that SN 2015bf share similarities with SNe~IIn-P. 
It was initially classified as a SN~IIn because of its narrow emission lines. 
The UV peak absolute magnitude of SN 2015bf ($M_{uvw2} \approx -20$\,mag) is also located at the faint end of SNe~IIn \citep{2014ApJ...787..157P}.
Moreover, SN 2015bf possibly suffered a flux drop before entering the radioactive decay phase.
Thus, given the massive CSM environment and small amount of synthesised nickel mass, the electron-capture-triggered explosion of a super-AGB star could be a possible scenario for SN 2015bf.

\section{Conclusions}\label{sec:conclu}

In this work, we present UV and optical observations of SN 2015bf. Our data cover the phase from +2\,d to +205\,d after the SN explosion. The light curves of SN 2015bf show a higher peak luminosity and a faster post-peak decline rate. Its light curve evolution within $\sim 60$ days is similar with SN 2013by. It is possible that SN 2015bf suffered a significant flux drop when entering the radioactive tail, but we can not conform this due to the lack of late-time photometric observations. 
The photospheric-phase spectra of SN 2015bf, with weak absorption of H$\alpha$ and slow velocity evolution of H$\beta$, are overall more similar to those of SNe~IIL. 

Moreover, observations of SN 2015bf enrich the SN sample with flash-ionised features at early phases. The spectrum {color{red} obtained at $t \approx 2$\,d} shows a blue continuum with strong FI emission lines, including ${\rm H\alpha}$, ${\rm H\beta}$, He {\sc ii}, and C {\sc iv} lines. Such FI features only lasted for about a week. Based on these features, one can speculate that the CSM was formed shortly before explosion at an enhanced mass-loss rate of $3.7 \times 10^{-3}\,{\rm M_\odot\,yr^{-1}}$. 
Assuming the tail of SN 2015bf is powered only by the radioactive decay of ${\rm ^{56}Co}$ and all the gamma rays are fully trapped, we estimated the nickel mass of SN 2015bf 
as $\sim 0.009 {\rm M_\odot}$ from its late-phase luminosity. 
Such a low nickel mass as well as dusty circumstellar environment, indicate that SN 2015bf could originate from an electron-capture explosion of a super-AGB star.  

\section*{Acknowledgements}

We acknowledge the support of the staff of the Lijiang 2.4\,m and Xinglong 2.16\,m/80\,cm telescopes.
This work is supported by the National Natural Science Foundation of China (NSFC grants 11633002, 12033002 and 11761141001) and the National Program on Key Research and Development Project (grant 2016YFA0400803). This work is also partially supported by the scholar Program of Beijing Academy of Science and Technology (DZ: BS202002) and the Strategic Priority Research Program of the Chinese Academy of Sciences, Grant No. XDB23040100. 
J. Zhang is supported by the NSFC (grant 11773067), by the Youth Innovation Promotion Association of the CAS (grant 2018081), and by the Ten Thousand Talents Program of Yunna for Top-notch Youth Talents. 
This work was partially supported by the Open  Project  Program  of  the  Key  Laboratory  of  Optical  Astronomy,  National Astronomical Observatories, Chinese Academy of Sciences. The LJT is jointly operated and administrated by Yunnan Observatories and the Center for Astronomical Mega-Science (CAS). Funding for the LJT has been provided by Chinese Academy of Sciences and the Peoples Government of Yunnan Province.

This work is based in part on observations with the twin Keck 10\,m telescopes on Maunakea, Hawaii. We are grateful to the staff at the Keck Observatory for their assistance, and we extend special thanks to those of Hawaiian ancestry on whose sacred mountain we are privileged to be guests. The W. M. Keck Observatory is operated as a scientific partnership among the California Institute of Technology, the University of California, and NASA; it was made possible by the generous financial support of the W. M. Keck Foundation. We thank S. Bradley Cenko and Daniel Perley for assistance with Keck spectral reductions. This work made use of {\it Swift}/UVOT data reduced by P. J. Brown and released in the {\it Swift} Optical/Ultraviolet Supernova Archive (SOUSA). SOUSA is supported by NASA's Astrophysics Data Analysis Program through grant NNX13AF35G.

A.V.F.'s supernova group is grateful for financial assistance from the TABASGO Foundation, the Christopher R. Redlich Fund, and the Miller Institute for Basic Research in Science (U.C. Berkeley). KAIT and its ongoing operation were made possible by donations from Sun Microsystems, Inc., the Hewlett-Packard Company, AutoScope Corporation, the Lick Observatory, the US National Science Foundation, the University of California, the Sylvia \& Jim Katzman Foundation, and the TABASGO Foundation. We thank the staff at Lick Observatory for their support. Research at Lick Observatory is partially supported by a generous gift from Google.

{\it Software:} IRAF \citep{1986SPIE..627..733T,1993ASPC...52..173T}, Zrutyphot(Mo et al. in prep.), SExtractor \citep{1996A&AS..117..393B}, DAOPHOT \citep{1987PASP...99..191S}

\section*{Data Availability}

The data underlying this article are available in the article and in its online supplementary material. We upload both the original spectra of SN 2015bf and the spectra used in Fig. \ref{fig:spectra} (corrected for redshift, host-galaxy extinction and Milky Way extinction, some spectra are also adjusted with the photometry) to the online supplementary material.




\bibliographystyle{mnras}
\bibliography{mnras_template} 




\appendix

\section{}

\begin{figure*}
\includegraphics[width=1.5\columnwidth]{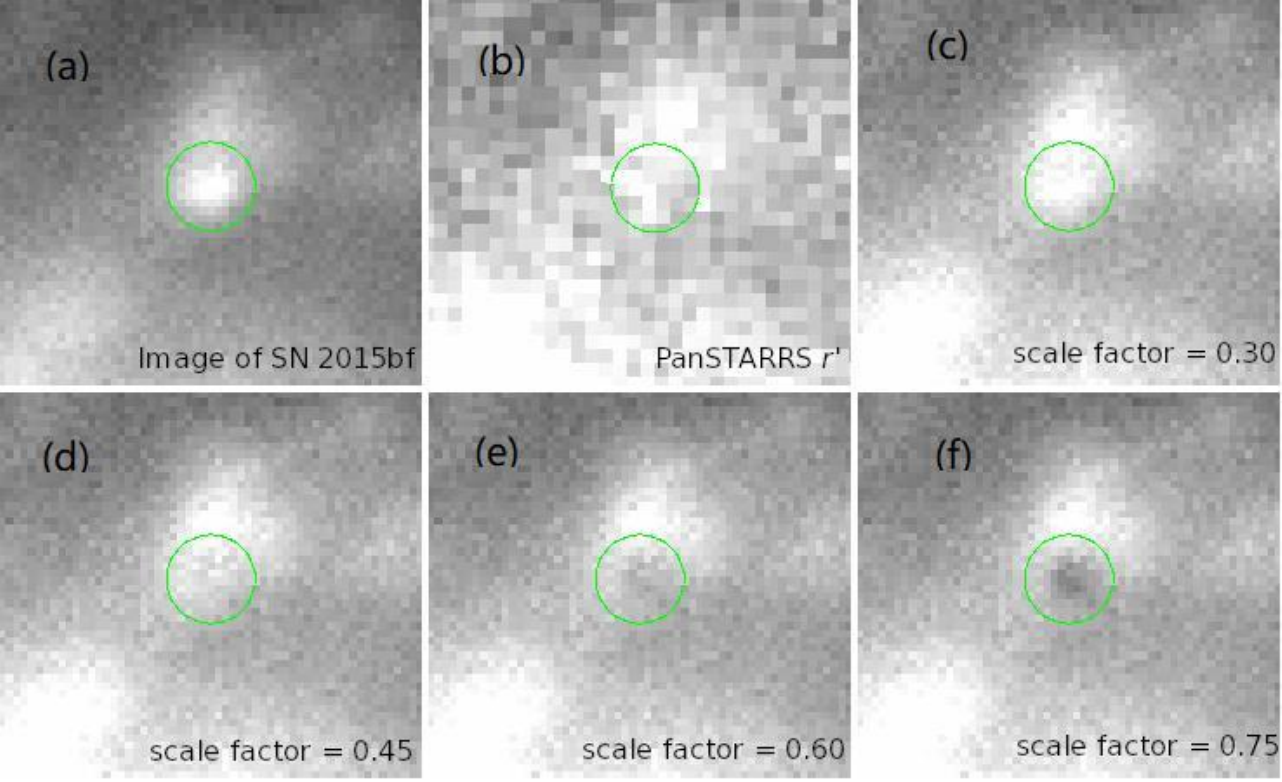}
\caption{(a): The $R$-band image of SN 2015bf obtained at $t \approx 205$\,d with LRIS on the Keck-I telescope. SN 2015bf is marked by a green circle. (b): $r^\prime$-band Pre-SN image from PS1. Panels (c), (d), (e) and (f) represent the excess/galaxy flux from which the SN flux has already been subtracted (created by multiplying the total flux in the PSF by a scale factor); see Section \ref{subsec:photometry}. 
After comparing with the flux at the SN location from the PS1 pre-SN image, we finally adopt a value of 0.45 for the scale factor as our final value.
We also try image subtraction technique by using the $r^\prime$-band image of PS1 (though the image of PS1 as template, and the resultant SN magnitude is 21.844 $\pm$ 0.086\,mag, which is $\sim$ 0.153\,mag fainter than the estimate by using a scale factor of 0.45.
\label{fig:keck_image}}
\end{figure*}

\begin{figure*}
\includegraphics[width=\columnwidth]{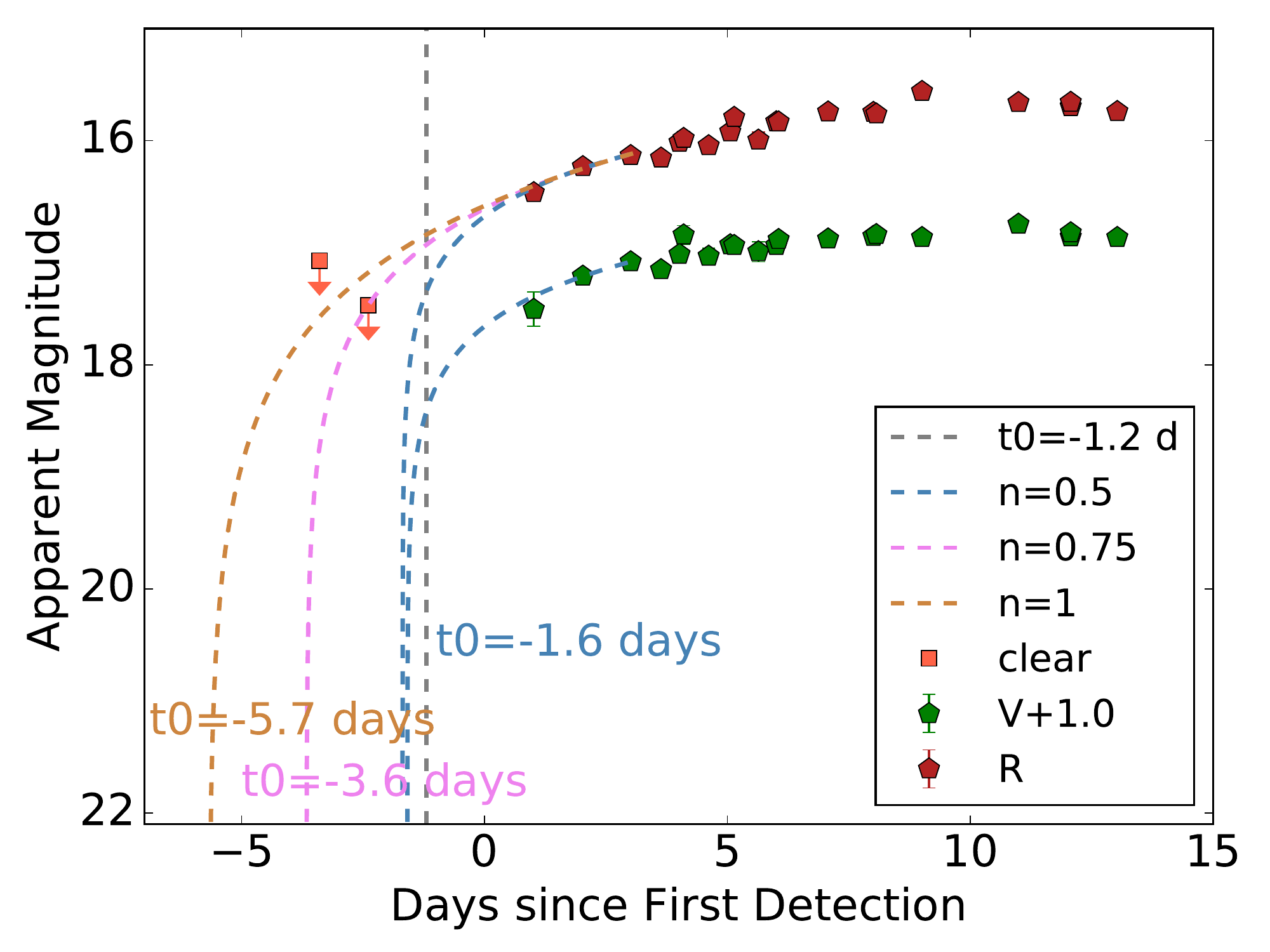}
\caption{${\rm A(t-t_0)^n}$ fit for the first three data points of SN 2015bf. The gray dashed line represents our estimated explosion time (the midpoint between the epoch of last nondection and discovery). The explosion time estimated by ${\rm n}=0.5$ is close to the midpoint. The expected light curve calculated by ${\rm n}=1$ (orange dashed line) is above the last nondetecion, while the expected light curve calculated by ${\rm n}=0.75$ (purple dashed line) is almost pass through the last nondection. 
Thus, we adopt ${\rm n}=0.75$ to estimate the earliest limit of explosion time.
\label{fig:powerlaw_fit}}
\end{figure*}

\begin{table*}
\centering
\caption{Photometric Standard Stars in the SN 2015bf Field}
\label{tab:standardstar}
\begin{tabular}{llcccccc} 
\hline
Num & $\alpha$(J2000) & $\delta$(J2000) & $U$ (mag) & $B$ (mag) & $V$ (mag) & $R$ (mag) & $I$ (mag) \\
\hline
1 & 23:24:51.50 & 15:13:02.8 & $\cdots$      & 14.738(0.005) & 14.108(0.002) & 13.751(0.002) & 13.326(0.002) \\
2 & 23:24:56.36 & 15:15:15.7 & $\cdots$      & 15.705(0.009) & 14.928(0.004) & 14.523(0.003) & 14.062(0.003) \\
3 & 23:24:35.85	& 15:13:28.2 & 16.462(0.008) & 16.406(0.018) & 15.873(0.008) & 15.486(0.006) & 15.077(0.006) \\
4 & 23:24:48.50  & 15:18:15.7 & $\cdots$      & 16.576(0.020) & 15.885(0.008) & 15.480(0.006) & 14.977(0.006) \\
5 & 23:24:43.96 & 15:13:38.1 & 17.272(0.010) & 17.143(0.032) & 16.459(0.014) & 16.037(0.010) & 15.627(0.010) \\
6 & 23:24:52.96 & 15:14:55.3 & 17.604(0.012) & 17.464(0.044) & 16.737(0.018) & 16.415(0.014) & 15.989(0.013) \\
7 & 23:24:36.14 & 15:15:45.5 & 17.713(0.014) & 16.958(0.027) & 16.078(0.010) & 15.513(0.007) & 14.981(0.007) \\
8 & 23:24:52.20 & 15:18:57.4 & $\cdots$      & 17.269(0.037) & 16.316(0.013) & 15.781(0.009) & 15.147(0.007) \\
9 & 23:25:03.56 & 15:21:16.5 & $\cdots$      & 13.265(0.002) & 12.739(0.001) & 12.417(0.001) & 12.031(0.001) \\
10 & 23:25:02.56 & 15:14:47.1 & 18.298(0.018) & 17.226(0.026) & 15.819(0.008) & 14.909(0.004) & 13.852(0.001) \\        
\hline
\end{tabular}
\end{table*}

\begin{table}
\centering
\caption{Clear-band Photometry of SN 2015bf}
\label{tab:clear}
\begin{tabular}{ccc} 
\hline
JD & Phase$^a$ (d) & Magnitude \\
\hline
2457372.60 & 4.83  & 16.51 (0.03)   \\
2457373.58 & 5.81  & 16.38 (0.03)  \\
2457374.60 & 6.83  & 16.40 (0.07)  \\
2457386.60 & 18.83 & 16.26 (0.03)  \\
2457399.60 & 31.83 & 16.45 (0.07)  \\
2457408.62 & 40.85 & 16.73 (0.04)  \\
2457415.62 & 47.85 & 16.86 (0.03)  \\     \hline
$^a$Phase relative to the estimated explosion date.
\end{tabular}
\end{table}

\begin{table*}
\centering
\caption{Optical Photometry of SN 2015bf}
\label{tab:photometry1}
\begin{tabular}{cccccccc} 
\hline
JD & Phase$^a$ (d) & $U$ (mag)  & $B$ (mag) & $V$ (mag) & $R$ (mag) & $I$ (mag) & Telescope\\
\hline
2457369.98 & 2.21  & $\cdots$      & $\cdots$      & 16.964(0.153) & 16.829(0.073) & 16.721(0.040) & LJT  \\
2457370.99 & 3.22  & $\cdots$      & 16.649(0.019) & 16.668(0.012) & 16.597(0.026) & 16.516(0.018) & TNT  \\
2457371.98 & 4.21  & $\cdots$      & 16.605(0.012) & 16.540(0.010) & 16.498(0.030) & 16.352(0.023) & TNT  \\
2457372.14 & 4.37  & 15.675(0.078) & 16.614(0.017) & $\cdots$      & 16.260(0.030) & $\cdots$      & LJT  \\
2457372.60 & 4.83  & $\cdots$      & 16.68(0.12)   & 16.61(0.06)   & 16.52(0.05)   & 16.32(0.06)   & KAIT \\
2457372.98 & 5.21  & $\cdots$      & 16.591(0.018) & 16.473(0.011) & 16.380(0.023) & 16.231(0.018) & TNT  \\
2457373.07 & 5.30  & 15.612(0.091) & 16.526(0.019) & 16.300(0.083) & 16.345(0.066) & 16.068(0.061) & LJT  \\
2457373.58 & 5.81  & $\cdots$      & 16.85(0.14)   & 16.49(0.07)   & 16.41(0.05)   & 16.26(0.06)   & KAIT \\
2457374.03 & 6.26  & $\cdots$      & 16.541(0.070) & 16.388(0.013) & 16.286(0.023) & 16.060(0.018) & TNT  \\
2457374.11 & 6.34  & 15.557(0.060) & $\cdots$      & 16.394(0.019) & 16.157(0.051) & 16.012(0.029) & LJT  \\
2457374.61 & 6.84  & $\cdots$      & 16.70(0.21)   & 16.45(0.09)   & 16.36(0.07)   & 16.21(0.09)   & KAIT \\
2457374.97 & 7.20  & $\cdots$      & 16.505(0.010) & 16.394(0.013) & 16.197(0.031) & 16.040(0.048) & TNT  \\
2457375.02 & 7.25  & 15.504(0.095) & 16.527(0.012) & 16.341(0.020) & 16.198(0.047) & 16.023(0.026) & LJT  \\
2457376.04 & 8.27  & $\cdots$      & 16.627(0.018) & 16.336(0.024) & 16.109(0.046) & 15.895(0.025) & LJT  \\
2457376.97 & 9.20  & $\cdots$      & 16.551(0.043) & 16.313(0.011) & 16.113(0.034) & 15.933(0.019) & TNT  \\
2457377.03 & 9.26  & $\cdots$      & 16.638(0.013) & 16.297(0.033) & 16.127(0.050) & 15.894(0.029) & LJT  \\
2457377.97 & 10.20 & $\cdots$      & 16.544(0.022) & 16.323(0.009) & 15.926(0.064) & 15.921(0.027) & TNT  \\
2457379.96 & 12.19 & $\cdots$      & 16.611(0.063) & 16.204(0.048) & 16.025(0.040) & 15.848(0.066) & TNT  \\
2457380.06 & 12.29 & 15.371(0.172) & $\cdots$      & $\cdots$      & $\cdots$      & $\cdots$      & LJT  \\
2457381.03 & 13.26 & $\cdots$      & 16.625(0.022) & 16.318(0.020) & 16.061(0.024) & 15.843(0.016) & TNT  \\
2457381.03 & 13.26 & 15.846(0.054) & 16.609(0.022) & 16.281(0.027) & 16.023(0.049) & 15.738(0.030) & LJT  \\
2457381.99 & 14.22 & $\cdots$      & 16.721(0.060) & 16.323(0.012) & 16.105(0.026) & 15.822(0.022) & TNT  \\
2457382.01 & 14.24 & $\cdots$      & $\cdots$      & $\cdots$      & 16.066(0.054) & 15.801(0.039) & LJT  \\
2457384.03 & 16.26 & 15.926(0.306) & 16.646(0.009) & 16.326(0.040) & 16.077(0.051) & 15.789(0.057) & LJT  \\
2457385.02 & 17.25 & $\cdots$      & 16.629(0.015) & 16.332(0.010) & 16.111(0.027) & 15.817(0.020) & TNT  \\
2457386.60 & 18.83 & $\cdots$      & 16.91(0.12)   & 16.45(0.05)   & 16.25(0.04)   & 15.95(0.05)   & KAIT \\
2457388.05 & 20.28 & $\cdots$      & 16.769(0.010) & 16.418(0.031) & 16.105(0.047) & 15.805(0.027) & LJT  \\
2457389.99 & 22.23 & $\cdots$      & 16.984(0.029) & 16.496(0.013) & 16.234(0.025) & 15.920(0.023) & TNT  \\
2457390.99 & 23.22 & $\cdots$      & 16.987(0.021) & 16.509(0.014) & 16.222(0.023) & 15.894(0.016) & TNT  \\
2457391.99 & 24.23 & $\cdots$      & $\cdots$      & 16.496(0.074) & 16.169(0.064) & 15.899(0.045) & LJT  \\
2457392.97 & 25.0  & $\cdots$      & 17.021(0.011) & 16.528(0.008) & 16.220(0.024) & 15.952(0.016) & TNT  \\
2457393.02 & 25.25 & 16.868(0.052) & 17.011(0.014) & 16.494(0.048) & 16.210(0.052) & 15.916(0.026) & LJT  \\
2457393.98 & 26.21 & $\cdots$      & 17.107(0.015) & 16.576(0.013) & 16.273(0.025) & 15.994(0.015) & TNT  \\
2457394.99 & 27.23 & $\cdots$      & 17.165(0.013) & 16.562(0.013) & 16.253(0.024) & 15.944(0.014) & TNT  \\
2457395.96 & 28.19 & $\cdots$      & 17.338(0.077) & 16.589(0.011) & 16.291(0.027) & 15.982(0.019) & TNT  \\
2457396.02 & 28.26 & 17.151(0.047) & 17.253(0.027) & 16.606(0.060) & 16.259(0.054) & 15.972(0.028) & LJT  \\
2457397.99 & 30.22 & $\cdots$      & 17.324(0.022) & 16.654(0.015) & 16.310(0.027) & 16.016(0.017) & TNT  \\
2457399.99 & 32.21 & $\cdots$      & 17.550(0.028) & 16.701(0.013) & 16.354(0.026) & 16.091(0.019) & TNT  \\
2457401.01 & 33.24 & $\cdots$      & 17.457(0.020) & 16.785(0.017) & 16.416(0.034) & 16.061(0.017) & TNT  \\
2457401.96 & 34.19 & $\cdots$      & 17.640(0.032) & 16.697(0.018) & 16.508(0.055) & 16.030(0.024) & TNT  \\
2457403.01 & 35.24 & 17.969(0.096) & 17.694(0.019) & 16.786(0.035) & 16.376(0.052) & 16.051(0.028) & LJT  \\
2457404.99 & 37.22 & $\cdots$      & 17.993(0.117) & 16.959(0.040) & 16.522(0.042) & 16.189(0.033) & TNT  \\
2457406.97 & 39.20 & $\cdots$      & 18.091(0.073) & 16.915(0.015) & 16.564(0.101) & 16.208(0.021) & TNT  \\
2457415.63 & 47.86 & $\cdots$      & 17.83(0.25)   & 17.24(0.13)   & 16.78(0.08)   & 16.45(0.08)   & KAIT \\
2457416.97 & 49.20 & $\cdots$      & 18.114(0.146) & 17.245(0.089) & 16.717(0.051) & 16.222(0.032) & TNT  \\
2457419.95 & 52.18 & $\cdots$      & 18.530(0.064) & 17.353(0.030) & 16.765(0.033) & 16.425(0.020) & TNT  \\
2457420.96 & 53.19 & $\cdots$      & 18.532(0.052) & 17.357(0.022) & 16.738(0.031) & 16.408(0.025) & TNT  \\
2457423.95 & 56.18 & $\cdots$      & 18.379(0.061) & 17.434(0.034) & 16.810(0.036) & 16.395(0.030) & TNT  \\
2457424.95 & 57.18 & $\cdots$      & 18.494(0.041) & 17.464(0.033) & 16.829(0.045) & 16.488(0.030) & TNT  \\
2457573.08 & 205.31 &$\cdots$ & $\cdots$ & $\cdots$ & 21.691(0.128) & $\cdots$ &Keck \\
\hline
$^a$Phase relative to the estimated explosion date.
\end{tabular}
\end{table*}

\begin{table}
\centering
\caption{{\it Swift} UVOT Photometry of SN 2015bf}
\label{tab:swift}
\begin{tabular}{cccc} 
\hline
JD & Phase$^a$ (d) & Filter & Magnitude \\
\hline
2457370.82 & 3.05  & $uvw2$ & 15.211(0.070) \\
2457372.20 & 4.43  & $uvw2$ & 15.411(0.078) \\
2457373.22 & 5.45  & $uvw2$ & 15.525(0.068) \\
2457377.18 & 9.41  & $uvw2$ & 16.574(0.077) \\
2457377.92 & 10.16 & $uvw2$ & 16.553(0.100) \\
2457379.10 & 11.33 & $uvw2$ & 17.008(0.085) \\
2457380.76 & 12.99 & $uvw2$ & 17.274(0.092) \\
2457383.32 & 15.55 & $uvw2$ & 17.926(0.119) \\
2457384.59 & 16.82 & $uvw2$ & 18.085(0.123) \\
2457387.08 & 19.31 & $uvw2$ & 18.480(0.160) \\
2457370.83 & 3.06  & $uvm2$ & 15.302(0.060) \\
2457373.22 & 5.45  & $uvm2$ & 15.503(0.062) \\
2457377.74 & 9.97  & $uvm2$ & 16.430(0.071) \\
2457379.11 & 11.34 & $uvm2$ & 16.787(0.078) \\
2457380.77 & 13.00 & $uvm2$ & 17.181(0.087) \\
2457383.33 & 15.56 & $uvm2$ & 17.911(0.123) \\
2457384.59 & 16.82 & $uvm2$ & 18.118(0.137) \\
2457387.09 & 19.32 & $uvm2$ & 18.988(0.245) \\
2457370.81 & 3.04  & $uvw1$ & 15.332(0.070) \\
2457372.22 & 4.45  & $uvw1$ & 15.347(0.065) \\
2457377.73 & 9.96  & $uvw1$ & 15.920(0.068) \\
2457379.10 & 11.33 & $uvw1$ & 16.174(0.071) \\
2457380.76 & 12.99 & $uvw1$ & 16.481(0.076) \\
2457383.32 & 15.55 & $uvw1$ & 16.924(0.089) \\
2457384.59 & 16.82 & $uvw1$ & 17.028(0.089) \\
2457387.08 & 19.31 & $uvw1$ & 17.619(0.122) \\
2457370.82 & 3.05  & $u$    & 15.492(0.072) \\
2457373.21 & 5.44  & $u$    & 15.407(0.065) \\
2457377.16 & 9.40  & $u$    & 15.602(0.060) \\
2457377.73 & 9.96  & $u$    & 15.567(0.067) \\
2457379.10 & 11.33 & $u$   & 15.653(0.067) \\
2457380.76 & 12.99 & $u$    & 15.788(0.069) \\
2457383.32 & 15.55 & $u$   & 16.015(0.073) \\
2457384.59 & 16.82 & $u$    & 16.135(0.072) \\
2457387.08 & 19.31 & $u$    & 16.344(0.079) \\
2457370.82 & 3.05  & $b$    & 16.709(0.084) \\
2457373.22 & 5.45  & $b$    & 16.673(0.073) \\
2457377.73 & 9.96  & $b$    & 16.632(0.072) \\
2457379.10 & 11.33 & $b$    & 16.572(0.070) \\
2457380.76 & 12.99 & $b$    & 16.600(0.071) \\
2457383.32 & 15.55 & $b$    & 16.679(0.073) \\
2457384.59 & 16.82 & $b$    & 16.826(0.074) \\
2457387.08 & 19.31 & $b$    & 16.829(0.077) \\
2457370.82 & 3.05  & $v$    & 16.876(0.138) \\
2457373.22 & 5.45  & $v$    & 16.473(0.094) \\
2457377.74 & 9.96  & $v$    & 16.220(0.084) \\
2457379.10 & 11.33 & $v$    & 16.242(0.083) \\
2457380.76 & 13.00 & $v$    & 16.332(0.087) \\
2457383.32 & 15.55 & $v$    & 16.398(0.091) \\
2457384.59 & 16.82 & $v$    & 16.410(0.088) \\
2457387.08 & 19.31 & $v$    & 16.439(0.095) \\       
\hline
$^a$Phase relative to the estimated explosion date.
\end{tabular}
\end{table}


\begin{table*}
\centering
\caption{Log of Spectroscopic Observations of SN 2015bf}
\label{tab:spectra}
\begin{tabular}{llccccc} 
\hline
UT Date & JD & Phase$^a$ (d) & Telescope & Instrument & Range (\AA) \\
\hline
2015/12/13     & 2457369.99 & 2.2   & LJT     & YFOSC+G3  & 3450--8830   \\
2015/12/16     & 2457373.08 & 5.3   & LJT     & YFOSC+G14  & 3760--7450   \\
2015/12/17  & 2457373.60 & 5.8   & Lick 3\,m    & Kast   & 3400--10,300  \\
2015/12/17     & 2457373.93 & 6.2   & Xinglong 2.16\,m & OMR  & 3700--8360   \\
2015/12/18     & 2457374.99 & 7.2   & LJT     & YFOSC+G3  & 3460--8830  \\
2015/12/19     & 2457376.01 & 8.2   & LJT     & YFOSC+G3  & 3460--8830   \\
2015/12/21     & 2457378.10 & 10.3  & LJT     & YFOSC+G3  & 3440--8830   \\
2015/12/31     & 2457388.05 & 20.3  & LJT     & YFOSC+G3  & 3450--8830   \\
2016/01/05     & 2457393.02 & 25.2  & LJT     & YFOSC+G3  & 3450--8830   \\
2016/01/07     & 2457394.50 & 26.7  & Keck-II  & DEIMOS & 4490--9450    \\
2016/01/08     & 2457396.03 & 28.3  & LJT     & YFOSC+G3  & 2440--8840   \\
2016/01/10     & 2457397.99 & 30.2  & Xinglong 2.16\,m & BFOSC  & 3710--8640 \\
2016/01/15     & 2457402.99 & 35.2  & LJT     & YFOSC+G3  & 3450--8840   \\
2016/01/30     & 2457418.01 & 50.2  & LJT     & YFOSC+G3  & 3440--8830   \\
2016/07/03     & 2457572.50 & 204.7 & Keck-I   & LRIS   & 3440--10,100  \\    
\hline
$^a$Phase relative to the estimated explosion date. 
\end{tabular}
\end{table*}


\bsp	
\label{lastpage}
\end{document}